\documentclass{jfm}
\usepackage{graphicx}
\usepackage{natbib}
\usepackage{xfrac}
\usepackage{comment}
\usepackage{upgreek}
\usepackage{siunitx}
\usepackage[a-1b]{pdfx} 
\usepackage{amsmath}
\allowdisplaybreaks
\usepackage{subcaption}
\usepackage{caption}
\usepackage{xcolor}
\usepackage{siunitx}

\let\realverbatim=\verbatim
\let\realendverbatim=\endverbatim
\renewcommand\verbatim{\par\addvspace{6pt plus 2pt minus 1pt}\realverbatim}
\renewcommand\endverbatim{\realendverbatim\addvspace{6pt plus 2pt minus 1pt}}
\makeatletter
\newcommand\verbsize{\@setfontsize\verbsize{10}\@xiipt}
\renewcommand\verbatim@font{\verbsize\normalfont\ttfamily}
\makeatother

\ifCUPmtlplainloaded \else
  \checkfont{eurm10}
  \iffontfound
    \IfFileExists{upmath.sty}
      {\typeout{^^JFound AMS Euler Roman fonts on the system,
                   using the 'upmath' package.^^J}%
       \usepackage{upmath}}
      {\typeout{^^JFound AMS Euler Roman fonts on the system, but you
                   do not seem to have the}%
       \typeout{'upmath' package installed. jfm.cls can take advantage
                 of these fonts,^^Jif you use 'upmath' package.^^J}%
      }
  \else
  \fi
\fi

\ifCUPmtlplainloaded \else
  \checkfont{msam10}
  \iffontfound
    \IfFileExists{amssymb.sty}
      {\typeout{^^JFound AMS Symbol fonts on the system, using the
                'amssymb' package.^^J}%
       \usepackage{amssymb}%
         \let\leq=\leqslant
         
      }{}
  \fi
\fi

\ifCUPmtlplainloaded \else
  \IfFileExists{amsbsy.sty}
    {\typeout{^^JFound the 'amsbsy' package on the system, using it.^^J}%
     \usepackage{amsbsy}}
    {\providecommand\boldsymbol[1]{\mbox{\boldmath $##1$}}}
\fi

\sbox{\astrutbox}{\rule[-5pt]{0pt}{20pt}}

\title[Journal of Fluid Mechanics]{\LaTeXe\ Input Guide for Authors}
\author{Cheng-Nian Xiao \and Inanc Senocak\corresp{\email{senocak@pitt.edu}}}

\affiliation{Department of Mechanical Engineering and Materials Science, \\ University
of Pittsburgh, 3700 O'Hara St, Pittsburgh, PA 15261, USA }

\date{10 August 2001 and in revised form 10 August 2004}
\pubyear{2001} 
\volume{000}
\pagerange{\pageref{firstpage}--\pageref{lastpage}}
\doi{S002211200100456X}

\begin{document}
 

 
\title{Speaker-wire vortices in stratified anabatic Prandtl slope flows and their secondary instabilities}
\maketitle
\begin{abstract}

Stationary longitudinal vortical rolls emerge in katabatic and anabatic Prandtl slope flows at shallow slopes as a result of an instability  when the imposed surface buoyancy flux relative to the background stratification is sufficiently large.
Here, we identify the self pairing of these longitudinal rolls as a unique flow structure. The topology of the counter-rotating vortex pair bears a striking resemblance to speaker-wires and their interaction with each other is a precursor to further destabilization and breakdown of the flow field into smaller structures. On its own, a speaker-wire vortex retains its unique topology without any vortex reconnection or breakup. For a fixed slope angle $\alpha=3^{\circ}$ and at a constant Prandtl number, we analyse the saturated state of speaker-wire vortices and perform a bi-global linear  stability analysis based on their stationary state. We establish the existence of both fundamental and subharmonic secondary instabilities depending on the circulation and transverse wavelength of the base state of speaker-wire vortices. The dominance of subharmonic  modes relative to the fundamental mode helps explain the relative stability  of a single vortex pair compared to the vortex dynamics in the  presence of two or an even number of pairs. These instability modes are essential for the bending and merging of multiple speaker-wire vortices, which  break up and lead to more dynamically unstable states, eventually paving the way for transition towards turbulence. This process is demonstrated via three-dimensional flow simulations with which we are able to track the nonlinear temporal evolution of these instabilities.


\end{abstract}

\section{Introduction}
 Prandtl model for katabatic and anabatic slope flows serve as a canonical model to comprehend the main dynamics of stably stratified flows around complex terrain, such as mountains and valleys \citep{prandtl1942, prandtl1952}. \textcolor{black}{The Prandtl model possesses  analytical one-dimensional laminar flow solutions which are exponentially damped sinusoidal velocity and buoyancy profiles along the slope-normal direction \citep{shapiro2004,fedorovich2009}.} The solution gives rise to a strong near-surface jet along the slope direction capped by a weak reverse flow as illustrated in figure \ref{fig:prandtlprofile}.
 In recent studies, we have investigated the linear stability of the Prandtl model for katabatic \citep{xiao2019} and anabatic \citep{xiao2020anabatic} slope flows, as well as the extended version of the Prandtl model with ambient winds \citep{xiao2020ambient} and discovered the existence of a stationary roll instability at small slope angles and traveling wave instability at steeper slopes. These studies have also helped establish the importance of the dimensionless stratification perturbation parameter in controlling the dynamics of stably stratified slope flows. 
 In the present investigation, we are interested in the progression of stationary roll instabilities in anabatic slope flows toward dynamically more unstable states. Despite accumulating experimental and numerical results in understanding of turbulence structure in neutral and unstable flows, the transition to turbulence under stable conditions still leaves open many significant  questions. 
 

Vortex pairing, which in its simplest form is a purely  two-dimensional dynamics where neighbouring vortices merge with each other, has been studied experimentally while observing
shear-layer growth\citep{winant1974vortex}. In related experiments,  \cite{miksad1972experiments} demonstrated that two-dimensional vortex structures are susceptible to three-dimensional instabilities. These three-
dimensional vortex instabilities are responsible for the creation of irregular flow at smaller scales which, however, do not significantly impact the large-scale vortex structures \citep{browand1976large}. The vortex instabilities are thus crucial for
the transition from two-dimensional to three-dimensional flows, and they continue to shape the features of the turbulent flow that develops further downstream. 
The simultaneous existence of stratification and bounding surface(s) is expected to further complicate the picture as outlined above. 
However, there has been scant theoretical approaches to adequately model vortex dynamics under these additional  flow conditions that are frequently encountered in real life.

For configurations without solid boundaries, \cite{crow1970stability} conducted a  stability analysis for a pair of parallel counter-rotating vortices in neutrally stable as well as quiescent ambient air and found sinusoidal symmetric bending of each vortex, which eventually led to the reconnection and vortex ring formation at locations where the distance between the neighboring vortices became minimal. \textcolor{black}{The linear stability of  counter-rotating vortex arrays in homogeneous fluid as introduced by \cite{mallier1993counter}  has been  analyzed in a numerical study limited to two spatial dimensions by \cite{dauxois1996stability}, which has been later extended to three dimensions by \cite{julien2002stability}. These works, however, only focused  on fundamental instabilities with the same wavelength as the base vortex array.  Three-dimensional \textcolor{black}{long-wave} instabilities of well-separated  vortex arrays have been studied by \cite{robinson1982stability}, which utilized the same approach  as \cite{crow1970stability} and focused on cooperative modes due to mutual induction only. This  analysis has been extended to study  vortex instability in rotating and stratified fluids by \cite{deloncle2011vortexarray}.}
 
Elliptic instabilities, which lead to anti-symmetric deformation of vortex cores were observed for vortices with non-circular streamlines \citep{pierrehumbert1986universal,hattori2021modal,schmid2012} and are caused by resonance of inertial vortex waves with the local  flow strain; however, they are suppressed by the presence of stable stratification, as shown by \cite{miyazaki1992three}. On the other hand, \cite{ledizes2009radiative} has shown that a single vertical columnar vortex can be unstable at stably stratified conditions due to the propagation and resonance of internal gravity waves, which is termed a radiative instability. \textcolor{black}{The dynamics of short-wavelength elliptic instability of co-rotating and counter-rotating vortex pairs in the presence of axial flows have been studied with DNS by \cite{schaeffer2008stability}, which found that axial flows tend to weaken the nonlinear dynamics of the elliptic instability.}

Another type of vortex instability that exists in rotating flows under stably stratified conditions is the so-called zig-zag  instability, which was first discovered and analysed by \cite{billant2000experimental,billant2010zigzag1}, and should not be confused with the zig-zag instabilities coined in \cite{clever1974transition} to describe three-dimensional instabilities of steady convection rolls.  The former type of zig-zag instabilities are caused by additional self induction as well as mutual induction among well-separated vortices due to a background density gradient which is aligned with the main columnar vortex, and they  lead to symmetric vortex bending in well-separated co-rotating vortex pairs and anti-symmetric bending in counter-rotating pairs.    

Besides studying the dynamics of the above-mentioned  model vortices that arise as approximate or exact solutions to the Navier-Stokes equations,  a second line of investigation has focused on stability analysis for roll structures that emerge as primary instability due to thermal convection or shear. \cite{corcos1984mixing} have identified that instabilities of vortices, which themselves arise as a primary instability in a one-dimensional shear layer, are responsible for vortex mergers and turbulence transition.
The stability of convection rolls that form on  flat as well as inclined heated surfaces in the absence of any stratification effects has been studied by \cite{clever1974transition,clever1977,busse1979instabilities}, which led to the discovery of the famous 'Busse balloon'  within the parameter space spanned by the horizontal wave number and Rayleigh number where these convection rolls are dynamically stable.   Similar stability analysis has been carried out for G\"ortler vortices under neutrally stratified conditions, and fundamental as well as subharmonic instability  modes have been discovered by \cite{hall1991linear, li1995goertler}. Like the results by \cite{pierrehumbert1982vortexinstab} for co-rotating Stuart vortices, \cite{li1995goertler} found that the fundamental modes cause a synchronous displacement of adjacent G\"ortler vortices, whereas the subharmonic modes lead to either merging or separation of two neighboring vortices.
All the research listed above has focused on the dynamics of elementary configurations with no more than two vortices, which are either counter-rotating (single vortex pair) or co-rotating (subharmonic mode of vortex array). However, the stability of stratified vortex configurations in the presence of a solid boundary involving more than two vortices, whether of the same sense of rotation or not, have found little attention so far.

In the present work, we expand our investigation of the Prandtl stratified slope flows with the stability analysis of the steady vortex pair arrays that arise as the saturated steady state of the primary instability under an anabatic heating condition at shallow slopes. \textcolor{black}{From now on, each counter-rotating vortex pair will be designated as a \textit{speaker-wire vortex}, and the motivation behind this terminology will be explained in more detail later.  }
There are several major differences between the instabilities of these speaker-wore vortices under the Prandtl's model and the other well-known vortex instabilities that we have mentioned above, such as Crow's instability, elliptic instability, zig-zag instability, secondary convection rolls and G\"oertler instabilities. Most importantly, the flow configuration described by Prandtl's model includes the following key components: Firstly, a constant vertical stable stratification which is at an oblique angle to the longitudinal rolls aligned with the streamwise direction and secondly,  a solid wall boundary  with its associated boundary layer of the Prandtl base flow. The simultaneous presence of all those features makes an analytical treatment very challenging and, to the best of our knowledge, is absent in the published  literature. As discovered in our earlier works \citep{xiao2019,xiao2020anabatic}, the primary roll instability of Prandtl's base flow consists of a pair of two counter-rotating vortices, hence any subharmonic secondary instability must involve at least two such pairs and four vortices in total, which is larger than the typical number  of vortices (1,2) in base flow configurations that have been investigated in the current literature on stability analysis.
Similar to the approach that we have pursued in ~\cite{xiao2019}, we apply linear bi-global stability analysis (LSA) to  identify the different  modes which can destabilize the base flow vortices. 
We further conduct analyses to determine how these modes may depend on external flow conditions and their role  in the turbulence transition of slope flows. 

\section{Properties of speaker-wire vortices}

\subsection{Governing Equations}
\begin{figure}
 \begin{subfigure}{0.44\textwidth}

	\centering
	\includegraphics[width=0.92\textwidth]{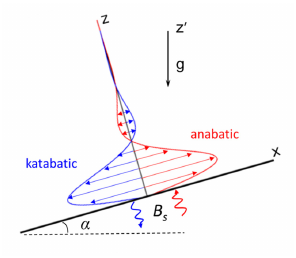}
   \caption{ }
	\label{subfig:sketchslope}
 \end{subfigure}
  \begin{subfigure}{0.54\textwidth}
	\centering
	\includegraphics[width=0.98\textwidth]{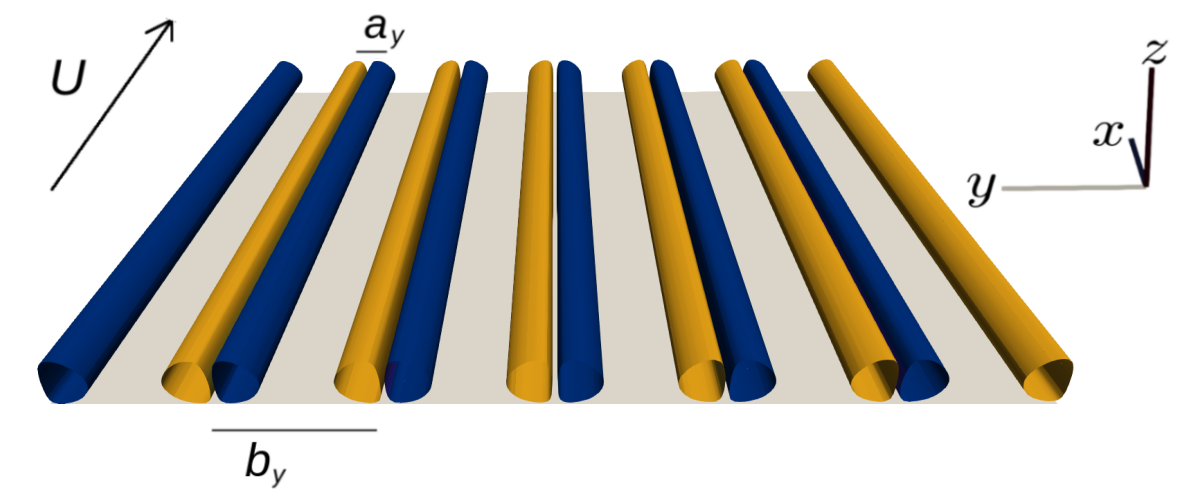}
   \caption{ }
	\label{subfig:slopevortices3d}
 \end{subfigure} \\
 \centering
   \begin{subfigure}{0.62\textwidth}
	\centering
	\includegraphics[width=0.98\textwidth]{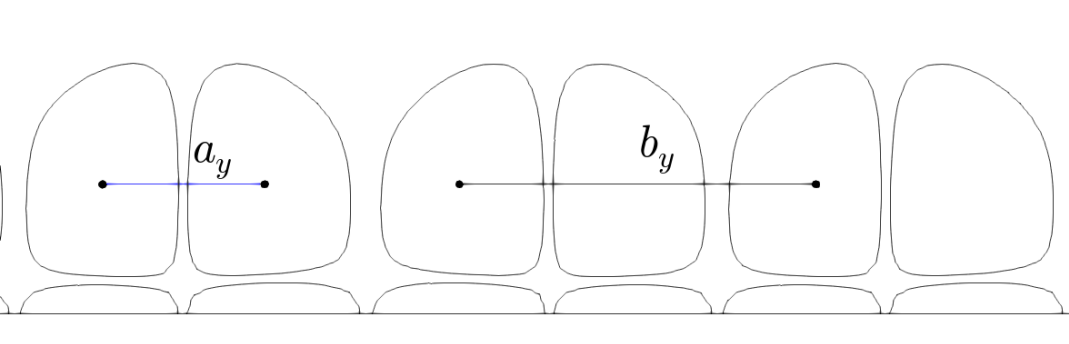}
   \caption{ }
	\label{subfig:slopevortices2d}
 \end{subfigure} 
 
	\caption{ Base flow profiles for slope flows under Prandtl's model: a) Sketch of slope flow geometry and the rotated coordinate system; b) Longitudinal vortex pairs (speaker-wire vortices) arising as instability of the laminar anabatic (upslope) Prandtl flow $\mathbf{U}$, with definition of coordinate axes,  roll separation $a_y$ and  speaker-wire vortex separation $b_y$ which is equivalent to the transverse wavelength  of the (primary) linear instability; c) 2D view of three speaker-wire vortex pairs with indication of the roll separation  $a_y$ and  vortex (pair) separation $b_y$.  }\label{fig:prandtlprofile}
\end{figure}

Let us consider the idealised Prandtl slope flow configuration as shown in figure \ref{subfig:sketchslope}, where $\alpha$ is the slope angle, and gravity $\mathbf{g}$ acts downward in the vertical direction. A constant positive buoyancy flux $B_S$ is imposed at the surface. We consider a rotated Cartesian coordinate system whose $x$ axis is aligned with the planar inclined surface. The direction normal to the slope surface is represented by the $z$ component of the position vector, whereas the cross-flow transverse direction is aligned by the $y$ coordinate. Let $u$ be the along-slope (longitudinal), $v$ be the cross-slope (transverse), and $w$ be the slope-normal velocity components, such that $\mathbf{u}=u_i=[u, v, w]$ is the velocity vector. The normalised gravity vector in the rotated coordinate system is then given by  $g_i=(g_1,g_2,g_3)=[\sin\alpha, 0, \cos\alpha]$.

The potential temperature, buoyancy, and the Brunt-V\"ais\"al\"a frequency are denoted by $\Theta,b, N$, respectively, where $N$, \textcolor{black}{which is assumed to be constant under Prandtl's model}, is related to the potential temperature as $N=\sqrt{\frac{g}{\Theta_r}\frac{\partial \Theta_e}{\partial z'}}$ and quantifies the background stratification, \textcolor{black}{where $z'$ denotes the vertical height  with respect to the horizontal.} The buoyancy is defined as a perturbation potential temperature as $b=g   (\Theta-\Theta_e)/ \Theta_r   $, where $\Theta_{r}$ is a reference potential temperature and $\Theta_e$ is the environmental (ambient) potential temperature. The kinematic viscosity and thermal diffusivity of the fluid  are denoted by $\nu,\beta$, respectively, and they are assumed to be constant. 
The transport equations for momentum with a Boussinesq approximation and buoyancy fields are written as follows:
\begin{align}
\frac{\partial \mathbf{u} }{\partial t}+\nabla \cdot(\mathbf{u} \otimes \mathbf{u})
=&~ -\frac{1}{\rho}\nabla  p + 
 b\mathbf{n}_{\alpha}+ \nu\Delta \mathbf{u} , \label{eqnslopemom}\\
\frac{\partial b}{\partial t} +  \nabla\cdot( b\mathbf{u})  = &~    \beta\Delta b   - N^2 \ (\mathbf{n}_{\alpha} \cdot \mathbf{u}) \label{eqnslopebuoy}.
\end{align}
where $\mathbf{n}_{\alpha}=(\sin\alpha,0,\cos \alpha)$ is the slope-normal unit vector.
The conservation of mass principle is imposed by a divergence-free velocity field:
\begin{align}
\nabla\cdot \mathbf{u}=0.\label{eqnslopecont}
\end{align}
 
\textcolor{black}{We impose  a positive buoyancy flux $B_{s}$ at the no-slip bottom surface, and impose periodic boundary conditions on the lateral boundaries. The top boundary satisfies the free-slip condition which means that the normal gradients of buoyancy and slope-normal as well as the transverse velocity components are zero, whereas the vertical velocity  component is set to zero. } \\ 
For the one-dimensional laminar flow problem,\cite{shapiro2004} extended the exact solution of \cite{prandtl1942} to include a constant surface flux instead of a constant buoyancy, \textcolor{black}{and}  the following characteristic flow scales have been introduced in \cite{fedorovich2009}:
\begin{align}
l_0 = &~ \text{Pr}^{\sfrac{-1}{4}}\ \nu ^{\sfrac{1}{2}} N^{\sfrac{-1}{2}}\sin^{\sfrac{-1}{2}}\alpha , \label{eqnlscale}\\
u_0 = &~ \text{Pr}^{\sfrac{1}{4}}\ \nu ^{\sfrac{-1}{2}}  N^{\sfrac{-3}{2}}B_{s} \sin^{\sfrac{-1}{2}}\alpha , \label{eqnuscale}\\
b_0 = &~ \text{Pr}^{\sfrac{3}{4}}\ \nu ^{\sfrac{-1}{2}}  N^{\sfrac{-1}{2}}B_{s} \sin^{\sfrac{-1}{2}}\alpha ,\label{eqnbscale}
\end{align}
where $\text{Pr}={\nu}/{\beta}$ is the Prandtl number. A time scale $t_0:=l_0/|u_0|=\sqrt{\nu\beta}N B_s^{-1}$ a shear scale $S_0:=|u_0|/l_0=\sqrt{Pr/\nu}N^{-1}B_s $ can also be defined from the above scales. We observe from (\ref{eqnlscale})-(\ref{eqnbscale}) that the length scale characterizing the laminar boundary layer thickness is independent of the surface flux $B_s$, whereas the magnitude of both the reference velocity and buoyancy scale varies linearly with $B_s$.
 \textcolor{black}{For all three-dimensional Navier-Stokes simulations, we use rectangular mesh elements and ensure in the discretization at least 2 points to resolve the length scale $l_0$ along both vertical and lateral directions.} Subsequently, these characteristic scales  will be applied to normalize all flow equations and quantities presented herein.
Specifically, the stratification perturbation number  $\Pi_s$ as introduced in \cite{xiao2019}  can be regarded as the imposed surface  buoyancy flux $B_s$ normalised by the background stratification scale $\beta N^2$. This unique parameter is determined from the given external flow parameters as follows: 
\begin{align}
\Pi_s \equiv \frac{B_s}{\beta N^2}.
\end{align}
\textcolor{black}{As can be seen from the previous work on slope flows by  \cite{fedorovich2009}, there exists a simple relation between the Reynolds number $Re$ based on the flow scales defined above and the  parameter $\Pi_s$, which is given by $Re=\Pi_s\cdot \sin(\alpha)/Pr$. We should note that a Reynolds number does not arise naturally in Prandtl slope flows because there is no external length or velocity scale imposed on the flow, whereas $\Pi_s$ arises naturally from an application of the Buckingham$-\pi$ theorem}
Furthermore, the significance of $\Pi_s$ is not confined to stratified slope flows only. In \cite{xiao2022impact}, we have shown that $\Pi_s$ emerges as an independent dimensionless parameter in open channel flows stratified simultaneously by a surface cooling flux and an independent ambient stratification.


\subsection{Longitudinal rolls and mode selection for the transverse wavelength}\label{sec:modeselect} 
The one-dimensional laminar Prandtl flow profile is susceptible to multiple types of linear instabilities when the surface buoyancy flux magnitude relative to the ambient stratification is sufficiently large, as characterised by the dimensionless number $\Pi_s=B_s/(\beta N^2)$(cf. \citealt{xiao2020anabatic}).
As shown in \cite{xiao2020anabatic}, at  shallow slopes such as $\alpha=3^{\circ}$, the most dominant linear instability is a stationary mode leading to longitudinal vortices aligned along the streamwise direction within the main upslope  flow. 
For each $\Pi_s$, there exists an optimal wavelength $\lambda_{\text{max}}(\Pi_s)$ that decreases with increasing $\Pi_s$ where the instability attains its maximal growth rate. Thus, at a specified $\Pi_s$ and for a given simulation domain with transverse length $Y$  which is an integer multiple of this optimal wavelength, i.e. $Y=n\lambda_{\text{max}}(\Pi_s), n\in \mathbb{N}$, the longitudinal rolls which arise  as  a result of linear instability have exactly this optimal transverse wavelength because modes of different wavelengths are out-competed due to their lower growth rates. Similarly, for any other arbitrary transverse domain length $Y$,  the emerging longitudinal roll instability has a wavelength $\lambda_y =\frac{1}{N} Y$, i.e. an integer fraction of the transverse length $Y$, and also possesses the maximal growth rate among all modes  which fits an integer number of times into $Y$.  Hence, by specifying the stratification perturbation $\Pi_s$ and the transverse domain size $Y$ in a simulation, we have pre-determined both the wavelength $\lambda_y$ of the emerging longitudinal rolls as well as the number $N$ of full rolls which are contained along the transverse extent. 
This is a technique we will make use of throughout this study to generate vortex pairs with a desired wavelength $\lambda_y$, which also equals the distance between to adjacent vortex pairs (see Figs. \ref{subfig:slopevortices3d}, \ref{subfig:slopevortices2d}).

\subsection{Speaker-wire vortex  structure at the saturated state}
The two-dimensional version of the slope flow equations (\ref{eqnslopemom})-(\ref{eqnslopecont})  where the along-slope direction is ignored, with imposed buoyancy flux $B_s$ at the no-slip bottom surface, can be solved via any time-stepping method to arrive at the steady vortices that arise as a saturated linear instability for slope angles less than $9^{\circ}$ when $\Pi_s$ is sufficiently large \citep{xiao2020anabatic}. To reproduce conditions of the local linear stability analysis presented in \cite{xiao2020anabatic}, the initial flow field is set to be the laminar Prandtl flow profile superposed with a weak sinusoidal disturbance varying along the transverse $y$ direction. The evolution of the flow field is tracked until a \textcolor{black}{quasi steady state} is reached, which produces  stationary  vortex structures as a result of nonlinear saturation of the growing initial disturbance at sufficiently large $\Pi_s$ value. 
\textcolor{blue}{We would like to  point out that  despite the existence of vortex instabilities of the saturated speaker-wire vortices as described below, these will only
be triggered when  very specific types of initial disturbances with the proper wavelengths such as those close to the strongest eigenmodes are already present within the flow field. In our simulations, we used regular rectangular
mesh elements which mostly suppress noisy numerical errors, thus allowing for the base vortices
to develop and saturate without being subject to vortex instabilities as described below.}

\begin{figure}
 \begin{subfigure}{0.48\textwidth}
	\centering
	\includegraphics[width=0.92\textwidth]{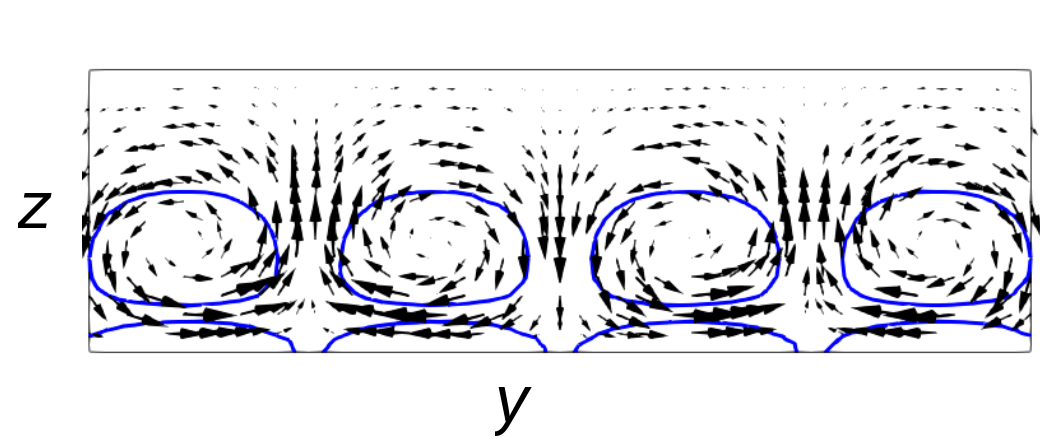}
   \caption{ }
	\label{subfig:vort_lin}
 \end{subfigure}
  \begin{subfigure}{0.48\textwidth}
	\centering
	\includegraphics[width=0.92\textwidth]{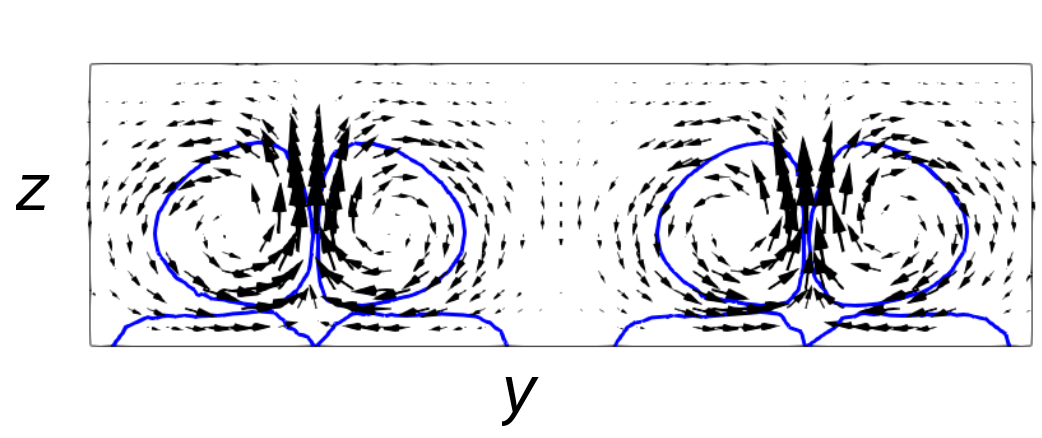}
   \caption{ }
	\label{subfig:vort_sat}
 \end{subfigure}
	\caption{ Speaker-wire vortices as a result of linear instability of the laminar anabatic Prandtl slope flow   at $\alpha=3^{\circ}$ with $\Pi_s=2.0$ and initial transverse wave number $k_y=1.22$: (a) during the linear growth phase of  initial harmonic disturbances and (b) steady state after nonlinear saturation. The blue lines are streamwise vorticity magnitude contours, whereas the black arrows represent the velocity field projected onto the transverse $yz$ plane. }\label{fig:anavortices_sat}
\end{figure}
A comparison between the vortex structure during the linear growth phase and the final steady state is shown in figure \ref{subfig:vort_lin} and \ref{subfig:vort_sat}. It is evident that even though the total number of vortex pairs (i.e. 2)  along the transverse direction has remained the same throughout, the originally uniformly distributed vortices have coalesced into one of their neighbors that has an opposite sense of rotation while moving away from their other neighbor, thus creating two counter-rotating pairs  that are clearly separated from each other. We will refer to these vortex pairs as a speaker-wire vortex system or more simply \textit{speaker-wire vortex} due to their striking resemblance to  actual speaker wires. The anti-parallel neighbor of a given vortex (longitudinal roll) within the same speaker-wire vortex (pair) will be called its sister roll.

A measure of the nonlinear modification of the original uniform vortex structure is the \textit{width-to-spacing ratio} $r=a_y/b_y$ between sister roll separation $a_y$ and speaker-wire vortex (pair) separation $b_y$,  as shown earlier in figure \ref{subfig:slopevortices3d}. For the initial disturbance varying harmonically  along the $y$ direction, $r=r_0=\frac{1}{2}$, and for the increased vortex separation as shown in figure \ref{subfig:vort_sat}, the value of $r$ has decreased below $r_0$. Clearly, the smaller the value of $r$, the more inhomogeneous the nonlinearly saturated vortices have become. This observation motivates the introduction of the \textit{vortex packing ratio}(VPR) defined as $\text{VPR}=2r=2a_y/b_y \times 100\% $, which measures how close two adjacent speaker-wire vortices  are from each other. During the linear growth phase, the rolls are distributed uniformly along the transverse direction with $r=1/2$, thus  VPR=100\% is maximal. When the rolls start to  coalesce into pairs to form speaker-wire vortices, VPR decreases due to clustering of two rolls within a pair. In the following, base speaker-wire vortices with VPR of 70\% or above will be regarded as \textit{tightly packed}, those with VPR between 50\% and 60\% are called \textit{intermediately packed}, whereas configurations whose VPR is less than 50\% are classified as \textit{loosely packed}.

Figure \ref{subfig:vort_a} shows the width-to-spacing ratio $r$ as a function of the transverse wavelength $\lambda_y$ of the primary roll instability, which equals the  vortex separation $b_y$, at constant $\Pi_s$. The  main vorticity magnitude  at the  vortex center  of the steady state longitudinal  rolls as a function of $\lambda_y$ is shown in figure \ref{subfig:vort_vortz} for different values of $\Pi_s$. It should be pointed out that since the longitudinal rolls are the manifestation of a primary instability, the roll separation $a_y$ and vortex (pair) separation  $b_y$  are a function of the transverse domain size which specifies the wavelength $\lambda_y$ of the emerging primary instability via the mode selection mechanism that we describe in Sec. \ref{sec:modeselect}. Thus, at a fixed  $\Pi_s$, the roll separation $a_y$,  vortex (pair) separation $b_y$ as well as the center vorticity are \textit{not} independent from each other; one can only specify the value for one of them by choosing the primary roll instability mode with the appropriate transverse wavelength $\lambda_y$, which then automatically fixes the other quantities of the emerging speaker-wire vortex structure. The dynamics of these speaker-wire vortices  is unlike the vortex instabilities presented in the works of \cite{pierrehumbert1982vortexinstab} and \cite{crow1970stability} in which the characteristics of the base flow vortices such as width and circulation can be freely specified and these vortices are not a product of a prior instability. For significantly larger transverse domain sizes as shown here,    the creation of an additional speaker-wire vortex will be favored, whereas for smaller transverse domain sizes, there is favorable pressure for two existing speaker-wire vortices to merge into one. Thus, 
the mode selection mechanism described earlier in Sec. \ref{sec:modeselect} 
aims to modify the number of flow vortices that fit within the specified transverse domain size to arrive at a  mode \textcolor{black}{with larger growth rate } at a different wavelength $\lambda_y$,  which is an integer fraction of the transverse domain extent. This can also be regarded as a two-dimensional instability of the speaker-wire vortices, which will be explained in more detail in the following sections. 

\section{Linear secondary instability analysis of speaker-wire vortices}
Let $(U,V,W,B)$ be the flow field in two-dimensional space of the steady longitudinal rolls,    and assuming that disturbances  to this base flow are waves of form \[\mathbf{q}(x,y,z,t) =\left[\hat{u}(y,z), \hat{v}(y,z), \hat{w}(y,z), \hat{p}(y,z),\hat{b}(y,z)\right]\exp({ik_x x + \omega t}),\], then the resulting linearised equations have the following form:
\begin{align}
ik_x\hat{u}+\frac{\partial \hat{v}}{\partial y}+ \frac{\partial \hat{w}}{\partial z} &= 0,\label{eqnslopelincont}\\
\omega \hat{u}+iUk_x\hat{u} +\frac{\partial U}{\partial y} \hat{v} + \frac{\partial U}{\partial z}\hat{w} & 
+\frac{\partial \hat{u}}{\partial y} V +  \frac{\partial \hat{u}}{\partial z}W \notag \\
&=  -ik_x\hat{p} -\frac{ Pr}{\Pi_s}\sin\alpha\left(-k_x^2\hat{u}+   \frac{\partial^2\hat{u}}{\partial y^2}+\frac{\partial^2\hat{u}}{\partial z^2} +\hat{b}\right),\\ 
\omega \hat{v}+iUk_x\hat{v}+\frac{\partial V}{\partial y}\hat{v}& +\frac{\partial V}{\partial z}\hat{w} 
+\frac{\partial \hat{v}}{\partial y}V+\frac{\partial \hat{v}}{\partial z}W  \notag \\
&= -\frac{\hat{p}}{\partial y}-\frac{Pr}{\Pi_s}\sin\alpha\left(-k_x^2\hat{v}+\frac{\partial^2\hat{v}}{\partial y^2} +\frac{\partial^2\hat{v}}{\partial z^2} \right),\\
\omega \hat{w}+iUk_x\hat{w} +\frac{\partial W}{\partial y}\hat{v}& +\frac{\partial W}{\partial z}\hat{w} +\frac{\partial \hat{w}}{\partial y}V+ \notag\\ \frac{\partial \hat{w}}{\partial z}W 
= -\frac{\partial\hat{p}}{\partial z} & -\frac{ Pr}{\Pi_s}\sin\alpha\left(-k_x^2\hat{w}+\frac{\partial^2\hat{w}}{\partial y^2} +\frac{\partial^2\hat{w}}{\partial z^2}+ \hat{b} \cot\alpha  \right),\\
\omega \hat{b}+iUk_x\hat{b}+\frac{\partial B}{\partial y}\hat{v}& +\frac{\partial B}{\partial z}\hat{w} +\frac{\partial \hat{b}}{\partial y}V+\frac{\partial \hat{b}}{\partial z}W  \notag \\
&= -\frac{\sin\alpha}{\Pi_s}\left(-k_x^2\hat{b}+\frac{\partial^2\hat{b}}{\partial y^2} +\frac{\partial^2\hat{b}}{\partial z^2} -( \hat{u}+\hat{w}\cot\alpha) \right), \label{eqnslopelinlast}
\end{align}
where $\hat{u},\hat{v},\hat{w},\hat{p},\hat{b}$ are  flow disturbances  varying along the slope normal and transverse directions normalised by the flow scales given in  (\ref{eqnlscale})-(\ref{eqnbscale}). The slope angle is fixed to $\alpha=3^{\circ}$, whereas  the normalised base flow field describing the steady vortices is denoted by $(U,V,W, B)$ and is obtained via the  simulation  procedure as described above.  \\

The linearised equations for bi-global stability analysis can be written as a generalised eigenvalue problem as follows:
\begin{align}
A(k_x)\mathbf{\hat{q}}(y,z)=\omega B(k_x)\mathbf{\hat{q}}(y,z),
\label{eig}
\end{align}
The two-dimensional complex disturbance  vector 
\[
\mathbf{\hat{q}}(y,z)=[\hat{u}(y,z),\hat{v}(y,z),\hat{w}(y,z),\hat{p}(y,z),\hat{b}(y,z
)]^T
\]
varies in the slope-normal (z) and transverse direction (y), where $(\hat{u},\hat{v},\hat{w})$ are the along-slope, cross-slope (transverse) and slope-normal disturbance velocity components.
As a bi-global stability analysis, the slope-normal and transverse dimensions are fully resolved and the disturbance variation along the streamwise direction is approximated by only one single Fourier mode with wave number $k_x$. When $k_x$ is zero, then the corresponding mode is two-dimensional without any streamwise variation, whereas a positive $k_x$ implies a full three-dimensional disturbance.  The appropriate boundary conditions for this problem are no slip for disturbance velocities at $z=0$ and $z\rightarrow \infty$, and for buoyancy disturbance $\partial \hat{b}/\partial z|_0=0, \hat{b}|_{z\rightarrow \infty}=0$ are imposed. The slope-normal derivative of pressure disturbance $\hat{p}$ is also set to zero at both $z=0$ and $z\rightarrow \infty$.  On both transverse boundaries $y=0,y=\lambda_y=b_y$, periodic conditions are imposed for all variables. The generalised eigenvalue problem (\ref{eig}) is discretised via  spectral elements in the transverse plane,
 which is available in Nektar++ \citep{nektar++}. For a base flow containing two full transverse spatial periods, i.e. two speaker-wire vortices, around 150 degrees of freedom are used for discretization in each transverse direction, and the resulting generalised eigenvalue problem is solved with the modified Arnoldi algorithm as implemented in Nektar++. Linear stability of the problem is associated with the real part of the eigenvalues $\omega$, where $\Re\{ \omega\}>0$ represents a positive exponential growth for the corresponding eigenmode, thus an unstable mode. The imaginary part of $\omega$
is the temporal oscillation frequency for the corresponding eigenmode, and $\Im\{ \omega\}=0$ represents a stationary mode. \textcolor{blue}{We have  conducted additional numerical studies to verify that the computed eigenvalues are robust  with respect to further mesh refinements by increasing  the order of each rectangle mesh element from 4 to 8, i.e. they change by less than half a per cent. We also ran eigenvalue analysis for the one-dimensional Prandtl base flow using the same vertical boundary conditions and were able to obtain the same numerical results as the separate in-house code which is described in \cite{xiao2019}.
This provides sufficient support that the computed eigenvalues are not of a spurious nature.}

\subsection{Instabilities of anabatic speaker-wire vortices at $3^{\circ}$ slope}


To investigate the secondary linear instability of speaker-wire vortices,  eigenvalues with the highest maximal real values  for a range of streamwise wave numbers $k_x$ are computed for different  vortex separation values $b_y$ and stratification perturbation parameter $\Pi_s$ of the base flow at the constant slope angle $\alpha = 3^{\circ}$.  We also assume a constant Prandtl number $Pr =0.71$ for all  cases. 

\begin{figure}
 \begin{subfigure}{0.48\textwidth}

	\centering
	\includegraphics[width=0.99\textwidth]{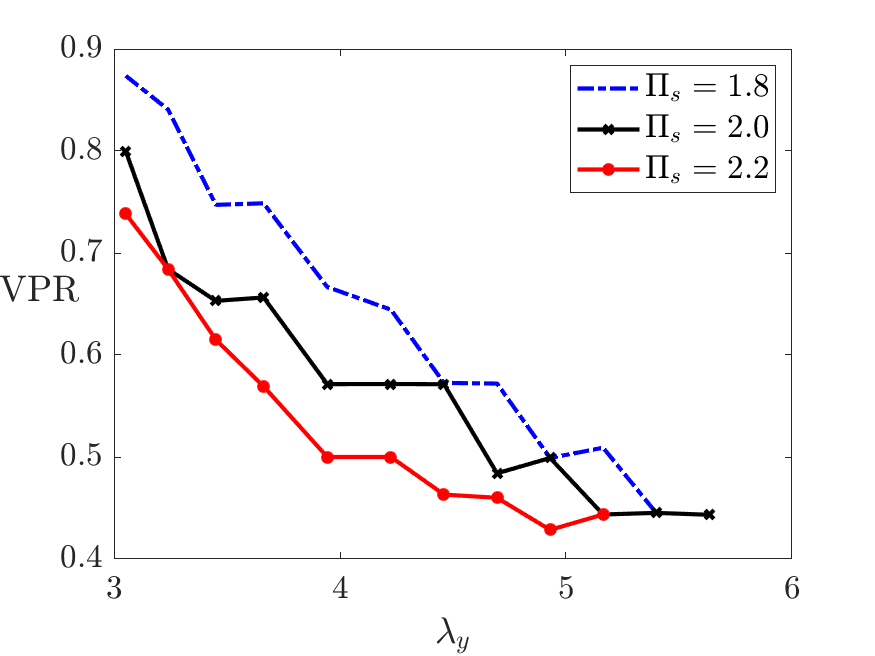}
   \caption{ }
	\label{subfig:vort_a}
 \end{subfigure}
  \begin{subfigure}{0.48\textwidth}
	\centering
	\includegraphics[width=0.99\textwidth]{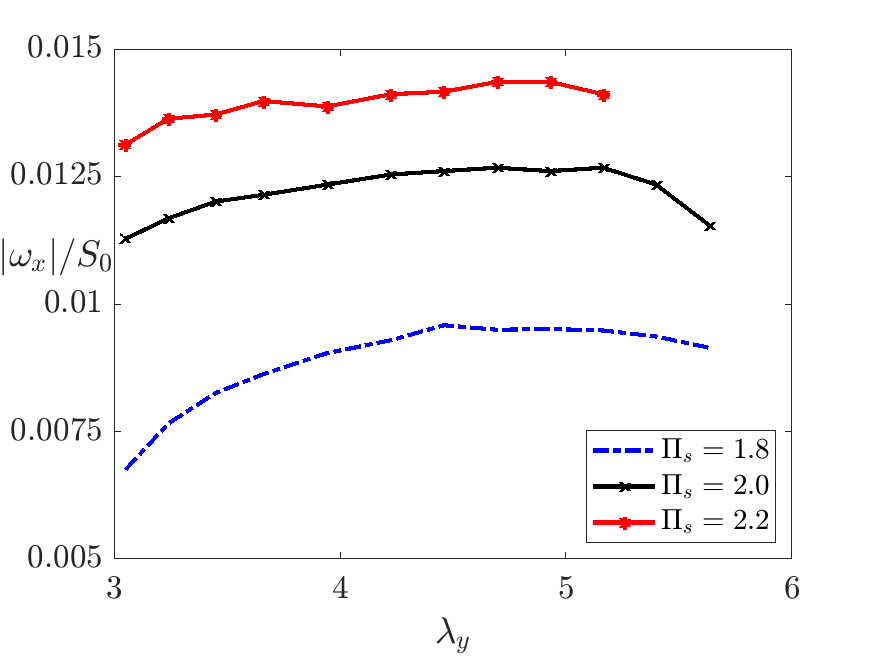}
   \caption{ }
	\label{subfig:vort_vortz}
 \end{subfigure}
 
	\caption{ Characteristics of  longitudinal vortex rolls for different values of $\Uppi_s$ after saturation from linear instability growth as a function of the  transverse wavelength $\lambda_y$ of the primary roll instability:  (a) vortex packing ratio(VPR), which is related to the width-to-spacing ratio $r$ via $r=a_y/b_y = \frac{1}{2}\text{VPR}$, and $b_y=\lambda_y$ is the  vortex separation;  (b) maximal streamwise vorticity magnitude  $|\omega_x|$ normalised by characteristic shear $S_0$}\label{fig:anavortices_lambda}
\end{figure}

We observe from figure \ref{fig:anavortices_lambda} that the  vortex separation  $b_y$ has a profound effect on the structure of the base flow when all other flow parameters are held constant, changing both the  circulation as well as vortex separation. Hence, we expect that the secondary instabilities arising from these base flow vortices will also be qualitatively different depending on $b_y$. In total, even at a constant slope angle and Prandtl number, three independent parameters determine the growth rates and oscillation frequencies of the secondary instability, which are the stratification parameter $\Pi_s$, the streamwise instability wave number $k_x=2\pi/\lambda_x$ and the transverse base flow wave number $k_y=2\pi/\lambda_y$, where $\lambda_y=b_y$ equals the  width of the base speaker-wire vortex. In the following, we will separately present and discuss the results of the stability analysis for base speaker-wire vortex  with different vortex separation values $b_y$.  \textcolor{black}{In our numerical investigations over many values for the separation $b_y$,  we have found that the qualitative behavior of the vortex dynamics can be classified  into  three categories, which can be  labeled as tightly packed, intermediately packed, and loosely packed.}\\

To illustrate the nonlinear effects of different vortex instabilities,  \textcolor{black}{the full set of Navier-Stokes Equations (\ref{eqnslopemom})-(\ref{eqnslopebuoy})  have been solved in two as well as three dimensions}  using nektar++5.0.0.  The same number of spectral elements DOF as for the bi-global linear stability analysis were used to resolve  the transverse $yz$ plane, and 16 Fourier modes were used for the along-slope direction $x$ in the three-dimensional simulations. The resulting animations   are available as supplementary movies.

\subsubsection{Case I: Tightly packed speaker-wire vortices}
\noindent \textbf{Subharmonic mode:} We begin our explorations by distinguishing between the subharmonic instabilities of the base vortices (i.e. those modes whose transverse wavelength spans twice the transverse wavelength of the primary vortex instability) and the fundamental instabilities with the same transverse wavelength as the primary mode. For  speaker-wire vortices with vortex (pair) separation $b_y =3.1$, we can see from figure \ref{subfig:vort_a} that the VPR  is over $70\%$. As figure \ref{subfig:vort_a} shows, the VPR  decreases with increasing vortex separation $b_y$ of the base speaker-wire vortices, which is equivalent to the primary transverse wavelength $\lambda_y$.   The base flow used for modal analysis consists of two speaker-wire vortices arising from the primary linear instability mode, i.e. the transverse domain size is twice the  wavelength of the primary vortex instability as described in \cite{xiao2020anabatic}.  

\begin{figure}
 \begin{subfigure}{0.48\textwidth}
	\centering
	\includegraphics[width=1\textwidth]{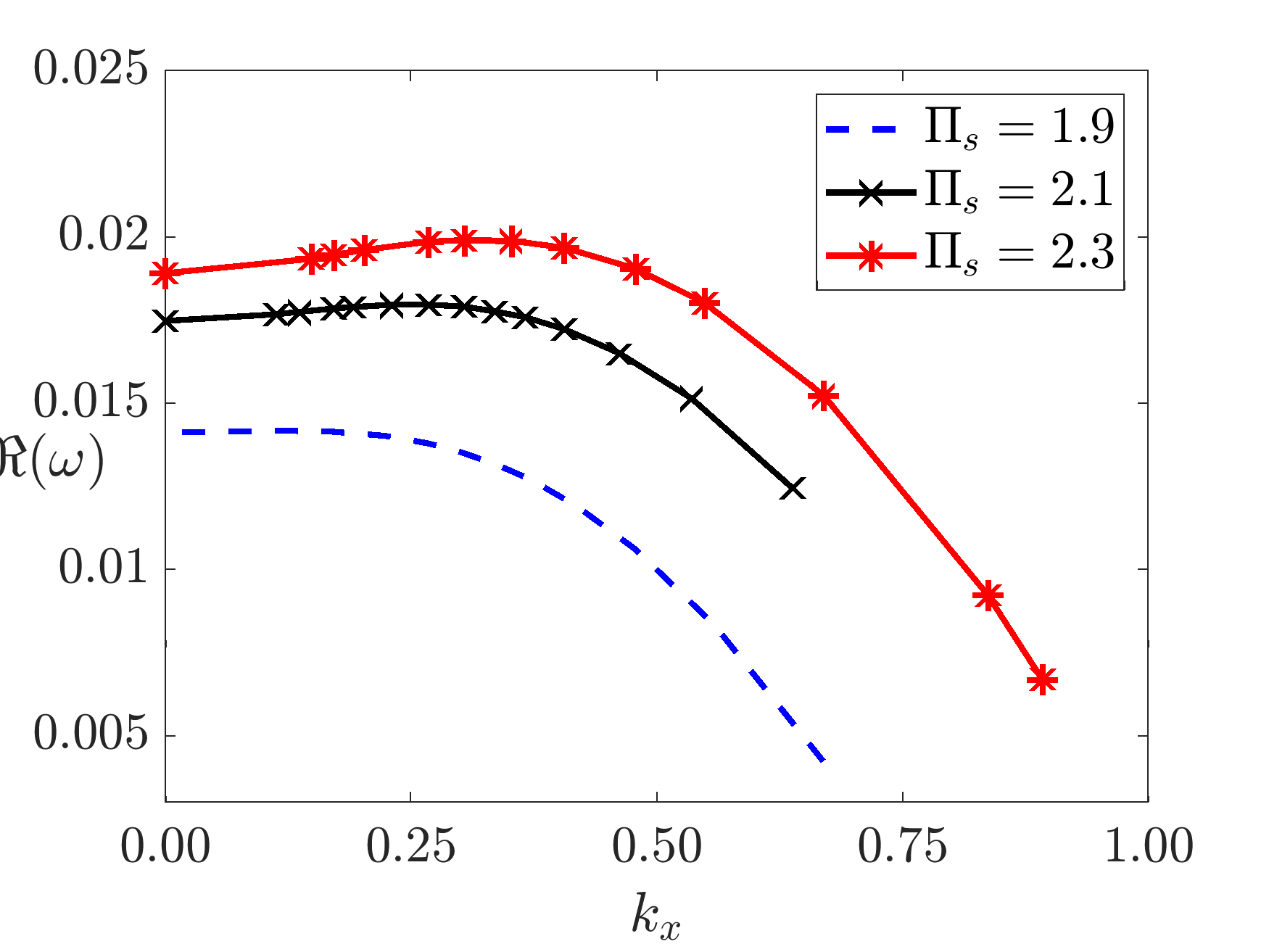}
   \caption{ }
	\label{subfig:subh06_gr}
 \end{subfigure}
  \hfill
  \begin{subfigure}{0.48\textwidth}
	\centering
	\includegraphics[width=1\textwidth]{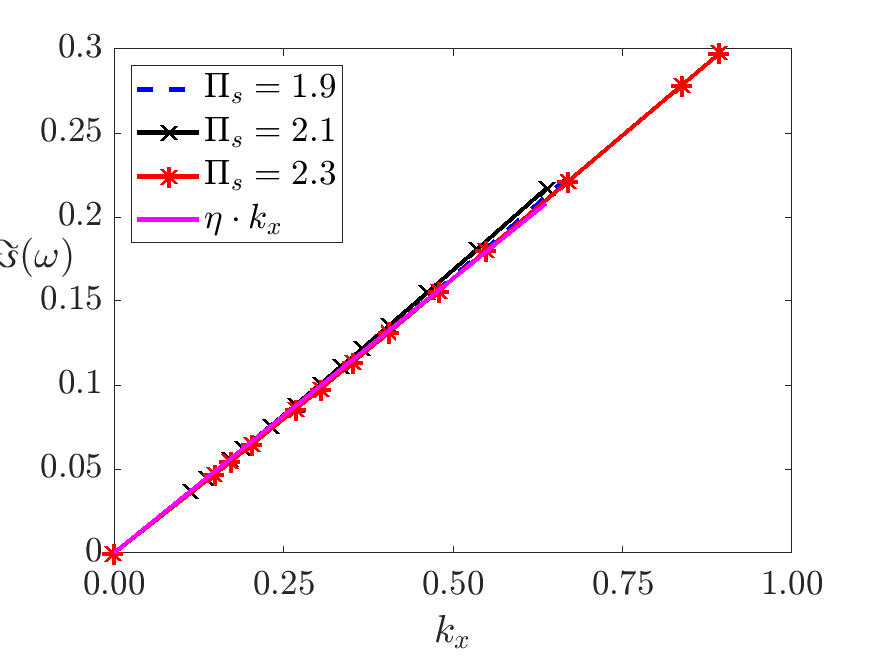}
   \caption{ }
	\label{subfig:subh06_osc}
 \end{subfigure}
 \vfill
 \vspace{15pt}
  \begin{subfigure}{0.59\textwidth}
	\includegraphics[width=0.87\textwidth]{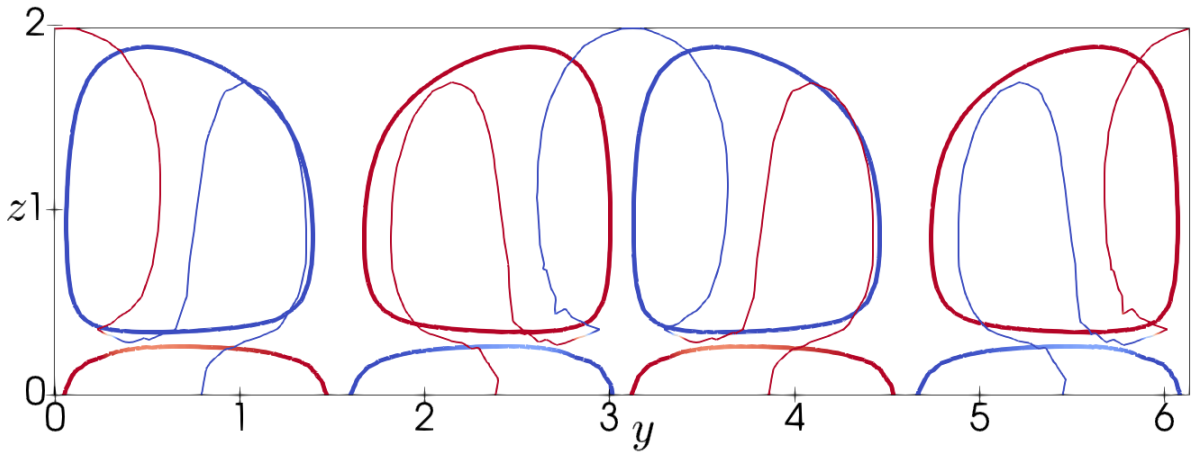}
   	\caption{}
	\label{subfig:subh06_2dcontour}
 \end{subfigure}
  \hfill
  \begin{subfigure}{0.4\textwidth}
	\includegraphics[width=0.7\textwidth]{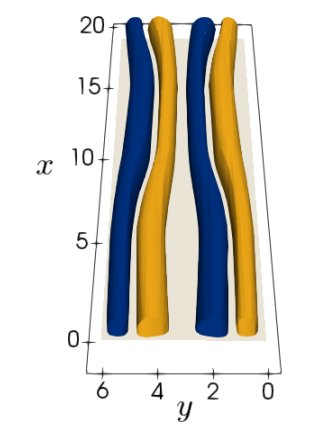}
   	\caption{}
	\label{subfig:subh06_3dcontour}
 \end{subfigure}
 \centering
   \begin{subfigure}{0.4\textwidth}
	\includegraphics[width=0.7\textwidth]{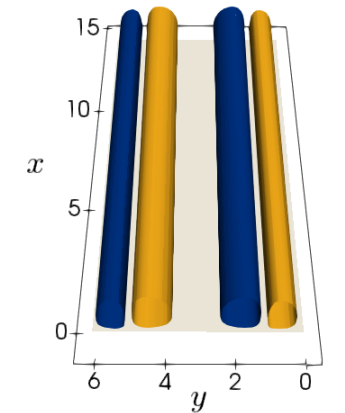}
   	\caption{}
	\label{subfig:subh062d_3dcontour}
 \end{subfigure}
	\caption{ Characteristics of  the subharmonic mode for tightly packed speaker-wire vortices at different values of $\Uppi_s$   as a function of the longitudinal wave number $k_x=2\pi/\lambda_x$  : (a) Growth rate $\Re(\omega)$;  (b) oscillation frequency $\Im(\omega)$ along with the linear  dispersion relation $\Im(\omega)=\eta\cdot k_x, \eta\approx 0.327$; (c) \textcolor{blue}{contours of streamwise vorticity  for the instability mode(thin lines) at $\Pi_s=1.9$ on the  $yz$ plane in relation to the base flow(thick lines);} (d) streamwise vorticity contours of  base speaker-wire vortices perturbed by three-dimensional subharmonic instability at $\Pi_s=2.3$;
	and (e) streamwise vorticity contours of  base speaker-wire vortices perturbed by two-dimensional subharmonic instability. \textcolor{blue}{All shown contours are at 5\% of the maximal vorticity magnitude.} Video of the evolution of the two-dimensional subharmonic instability for tightly-packed vortices is available  as supplementary Movie 1.
	}
\end{figure}

The  growth rates of the most unstable secondary modes for streamwise or longitudinal wave numbers  $k_x$ within the range $[0,1.0]$ are shown for three different values of $\Pi_s =1.9,2.1,2.3$ in figure \ref{subfig:subh06_gr}. As expected, the growth rate of any mode at a fixed streamwise wave number $k_x$ grows with increasing $\Pi_s$ due to an increase in the surface heat flux. 
It turns out that all the secondary instability  modes with the largest eigenvalues have a transverse wavelength which is exactly equal to the total transverse extent of the base flow, i.e. twice the width $b_y$ between adjacent base flow speaker-wire vortices, as can be seen in the 2D transverse contour plot for the streamwise vorticity $\omega_x$ shown in figure \ref{subfig:subh06_2dcontour}, suggesting that they are indeed \textit{subharmonic} instabilities.  
Two-dimensional modes  have zero longitudinal wave numbers, i.e. $k_x=0$, and thus are constant along the streamwise direction.
For the three-dimensional modes with $k_x>0$, we can see from Figure \ref{subfig:subh06_gr} that the growth rates  
 reaches a maximal value slightly exceeding the 2D  growth rate at a optimal wave number $k_x\approx 0.4$, and from that point on they remain nearly constant to decrease very slowly to the 2D growth rate at $k_x=0$. 
Due to the fact that  there is very little variation in growth rate as a function of the streamwise wave number for values less than $k_x\approx 0.4$ corresponding to the strongest mode, we can conclude that extremely long-wave or even two-dimensional structures which are known to appear in  stably stratified channel flows \citep{garcia2011turbulence} can manifest themselves when the base vortices are destabilized.
All 3D modes with positive longitudinal wave numbers $k_x>0$ are oscillatory whose frequency increases monotonically with growing $k_x$, as shown in figure \ref{subfig:subh06_osc}. The two-dimensional mode with $k_x=0$  is  stationary, i.e. with  zero imaginary part of its eigenvalue. At small wave numbers $k_x$, the oscillation frequencies of the three-dimensional mode asymptotically decay to zero with decreasing $k_x$ to converge to the stationary two-dimensional mode at $k_x=0$. As figure \ref{subfig:subh06_osc} shows, the  normalised frequencies of all three cases shown here appear to obey a simple linear dispersion relation given by $\Im(\omega)=\eta\cdot k_x$, where  $\eta \approx 0.327$  is empirically determined to fit all three curves. The accuracy of this fit shows that the group speed  of the strongest subharmonic vortex instabilities given by $c=\frac{\partial \Im(\omega)}{\partial k_x}=\eta$ is nearly constant  and equals $\eta$.

Existence of  both two-dimensional  and three-dimensional subharmonic vortex instabilities has also been identified in the study of co-rotating Stuart vortex arrays  by \cite{pierrehumbert1982vortexinstab} or as secondary instabilities in a shear layer by \cite{corcos1984mixing}, where they are shown to be responsible for the merging of neighboring vortices. Similarly, as visualised in  figure \ref{subfig:subh06_3dcontour}, the three-dimensional secondary subharmonic modes of the speaker-wire vortices act  by bending adjacent speaker-wire vortices in opposite directions  as to facilitate their reconnection. As shown in figure \ref{subfig:subh062d_3dcontour}, the effect of the two-dimensional subharmonic mode is to strengthen one roll while weakening its sister roll within the same speaker-wire vortex, with the aim of  halving the total  number of vortices present within the domain. Depending on the range of permissible  streamwise wavelength, which itself depends on the length of the domain, the strongest subharmonic mode is either purely two-dimensional or three-dimensional.  When the streamwise domain length is $L_x$, a mode with wave number within $2\pi/L_x$ of the optimal wave number $k_O$ will  manifest itself.  Thus, for large values of $L_x$, a three-dimensional mode with wavelength close to the optimal wavelength will be favored; however, in the case when the streamwise domain length is less than $2\pi k_O$, the two-dimensional subharmonic mode will come out the strongest.\\

\noindent \textbf{Fundamental mode:}  
 The fundamental secondary instability has, by definition, a transverse wavelength equaling the transverse wavelength $\lambda_y$ of the primary instability, or one half the wavelength of a subharmonic mode. Thus, 
the base flow used in the present analysis consists of only one single speaker-wire vortex that rises from the primary linear instability mode in order to ensure that the maximal transverse wavelength $\lambda_y$ of the eigenmodes computed here can not exceed the transverse wavelength of the primary vortex instability. 

\begin{figure}
 \begin{subfigure}{0.48\textwidth}
	\centering
	\includegraphics[width=1\textwidth]{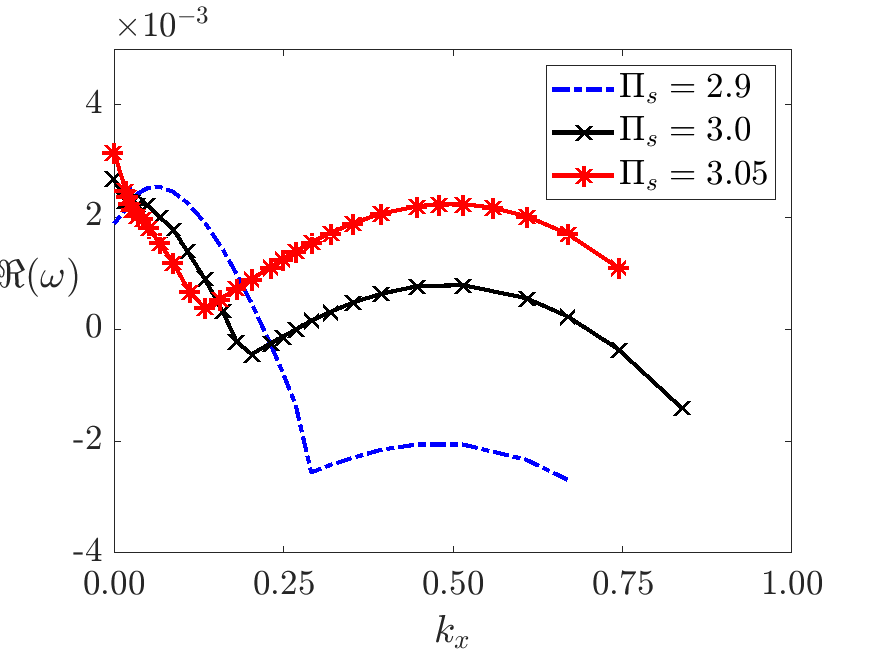}
   \caption{ }
	\label{subfig:fund03_gr}
 \end{subfigure}
  \hfill
  \begin{subfigure}{0.48\textwidth}
	\centering
	\includegraphics[width=1\textwidth]{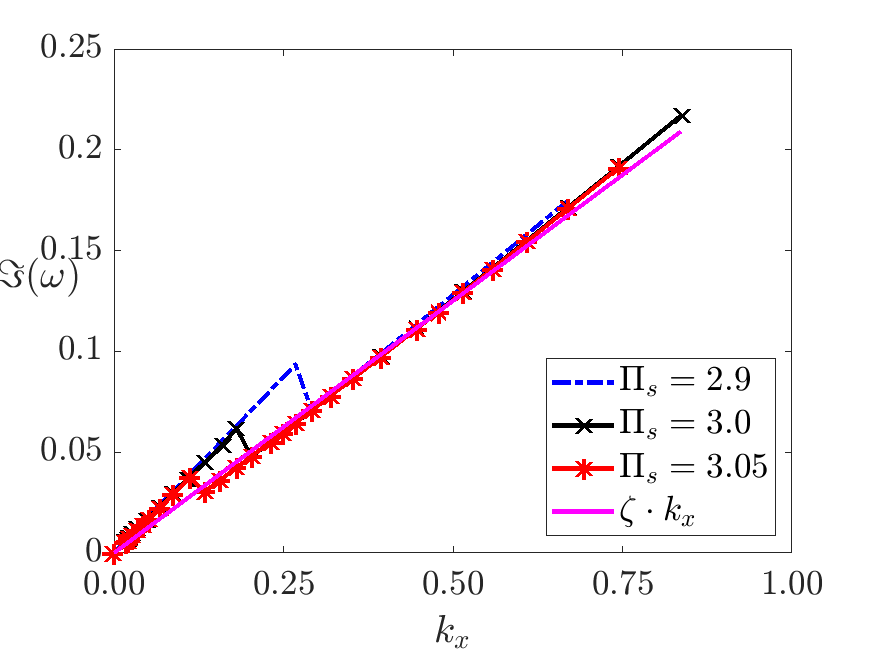}
   \caption{ }
	\label{subfig:fund03_osc}
 \end{subfigure}
 \vfill
 \vspace{15pt}
   \begin{subfigure}{0.48\textwidth}
		\centering
	\includegraphics[width=0.8\textwidth]{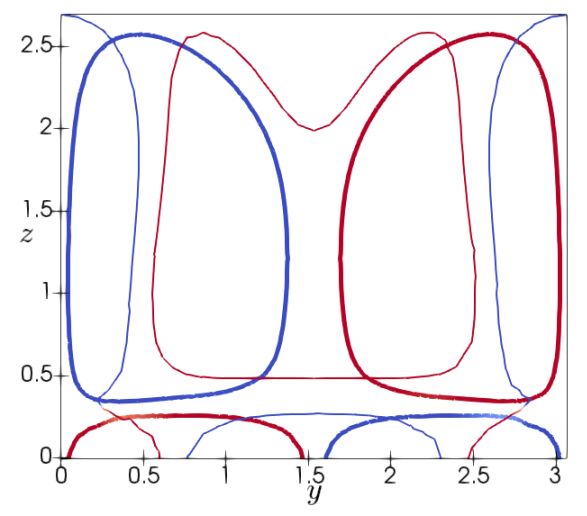}
   	\caption{}
	\label{subfig:fund03_contour2d}
 \end{subfigure}
  \hfill
      \begin{subfigure}{0.4\textwidth}
      	\centering
	\includegraphics[width=0.6\textwidth]{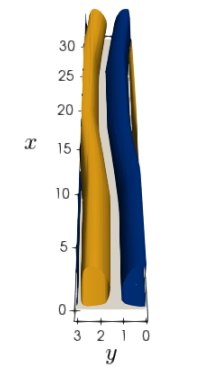}
   	\caption{}
	\label{subfig:fund03_contour3d}
 \end{subfigure}
 
	\caption{ Characteristics of  the fundamental mode  for tightly packed speaker-wire vortices  at different values of $\Uppi_s$ as a function of the longitudinal wave number $k_x$: (a) Growth rate $\Re(\omega)$;   (b) oscillation frequency $\Im(\omega)$ along with the linear dispersion relation $\Im(\omega)=\zeta\cdot k_x, \zeta\approx 0.25$; (c)  streamwise vorticity contours of  two adjacent rolls from  neighboring speaker-wire vortices at $\Pi_s=3.05$     on the transverse $yz$ plane and (d)  streamwise vorticity contours of  base speaker-wire vortices perturbed by the fundamental instability at $\Pi_s=3.05$. \textcolor{blue}{All contours are displayed at 5\% of the maximal vorticity magnitude.}   } 
\end{figure}

The  growth rates and oscillation frequencies of the most unstable  modes for streamwise wave numbers  $k_x$ within the range $[0,1.0]$ are shown for three different values of $\Pi_s = 2.9,3.0, 3.05$ in figure \ref{subfig:fund03_gr} and figure \ref{subfig:fund03_osc}. As shown in figure \ref{subfig:fund03_contour2d}, the transverse wavelength of the fundamental mode is approximately 3, thus equal to the base vortex separation $b_y$.  Compared with the subharmonic mode that we discussed previously and visualised in figure \ref{subfig:subh06_gr}, the growth rates of these fundamental modes are around five times smaller despite  possessing  larger normalised surface fluxes, which is also indicated by the larger $\Pi_s$ values in their base flows.
At the lower values of $\Pi_s$ as displayed for the subharmonic modes in figure \ref{subfig:subh06_gr}, there exists no unstable fundamental modes. This signifies that there is a clear separation between the fundamental and subharmonic modes, and that the fundamental modes are inherently weaker than their subharmonic counterparts, whose transverse wavelength is twice as large. The flow-physical implication of this clear gap between the subharmonic and fundamental modes is that the most dominant vortex dynamics is the merger or reconnection of two adjacent speaker-wire vortices involving \textit{four} rolls in total, rather than reconnection within a  single vortex pair.

Figure \ref{subfig:fund03_gr} shows that with  increasing  normalised surface heat flux $\Pi_s$, the growth rate of the strongest mode at the large as well as smallest streamwise wave numbers $k_x$  also grows. However, for intermediate wave number values   $0.05<k_x<0.2$, the growth rates  are higher for lower $\Pi_s$. 
Similar to what we have observed in the case of subharmonic modes, there exists stationary two-dimensional fundamental modes that do not vary along the streamwise along-slope direction; their growth rates are shown in figure \ref{subfig:fund03_gr}  as the value at the wave number $k_x=0$.
In contrast to the subharmonic modes, the strongest three-dimensional fundamental modes ($k_x>0$)  have lower growth rates than their two-dimensional counterparts for all values of $k_x$.  

All the three-dimensional fundamental modes are oscillatory, and their frequencies increase monotonically with the streamwise wave number $k_x$, as shown in figure \ref{subfig:fund03_osc}. At small  wave numbers $k_x$, the frequency of the three-dimensional fundamental mode asymptotically decays to zero to  become the stationary two-dimensional mode at $k_x=0$. It can be seen that for larger wave numbers with $k_x>0.35$, the  normalised frequencies of the strongest fundamental modes at the three values  for $\Pi_s$ shown here approximately fit a simple linear dispersion relation given by $\Im(\omega)=\zeta\cdot k_x$, where  $\zeta \approx 0.25$  is empirically determined to fit all three curves. At lower wave numbers $k_x<0.2$, the  fundamental mode  from another branch of the spectrum with higher frequency becomes stronger than the dominant mode at larger wave numbers, which manifests itself in an abrupt jump of the oscillation frequency $\Im(\omega)$ 
 of the strongest fundamental mode as a function of $k_x$ as shown in figure \ref{subfig:fund03_osc}.

Fundamental instabilities of longitudinal rolls, i.e. modes which have the same transverse wavelength  as the base vortices, have been extensively studied in other previous works as well. The most prominent representative of such a mode is the  Crow instability for a pair of vortices suspended in unstratified air \citep{crow1970stability}.  Fundamental modes have also been identified as instabilities of    Rayleigh-Benard convection rolls  which aim at distorting the structure and spacing of the rolls to bring them closer to the optimal wavelength \citep{clever1974transition}. 
More recently, fundamental vortex instabilities have also been found in strongly stratified fluids such as the so-called zig-zag instability studied in \cite{billant2010zigzag1}. Similar to the fundamental vortex instabilities discovered in \cite{pierrehumbert1982vortexinstab}, we  expect that the main effect of the fundamental modes is to cause parallel displacement and symmetric distortion of all  vortices.
As visualised in \ref{subfig:fund03_contour3d}, the three-dimensional fundamental mode  causes  sinusoidal bending and distortion of each speaker-wire vortex. However, unlike the well-known symmetric Crow  instability \citep{crow1970stability} which bends two sister rolls toward each other to facilitate their connection, the fundamental mode for speaker-wire vortices can only bend both vortices within a pair along the same direction, as shown in figure \ref{subfig:fund03_contour3d}.\\

\noindent \textbf{Coherent nature of single speaker-wire vortex } 
We observe from figure \ref{subfig:subh06_3dcontour} and figure \ref{subfig:fund03_contour3d}, in both subharmonic and fundamental modes, the two rolls within the same speaker-wire vortex always bend in the same direction, thus preventing a vortex reconnection or merger. This means that a single pair of vortices can remain in its basic pair structure even after the initial onset of instabilities, thus justifying their designation as an unique coherent vortex structure. Similar vortex structures which remain stable and coherent over large wavelengths have been observed in the \textit{Langmuir vortices} on the surface of the seas and oceans as described in \cite{craik_leibovich1976langmuir}.\\ 

\subsubsection{Case II: Intermediately packed speaker-wire vortices}

\noindent \textbf{Subharmonic and fundamental modes:} For a larger  speaker-wire vortex separation   of $b_y =4.1$, we observe from figure \ref{fig:anavortices_lambda} that the VPR of speaker-wire vortices is decreasing to around 60\%. This indicates that relative to the previous vortex configuration (i.e. Case I) with a smaller transverse vortex wavelength, the two speaker-wire vortices   move tighter towards each other after reaching the end of their linear growth phase, and leading to a clearer separation of adjacent speaker-wire vortices.  

\begin{figure}
\begin{subfigure}{0.49\textwidth}
\centering
    \includegraphics[width=1.0\textwidth]{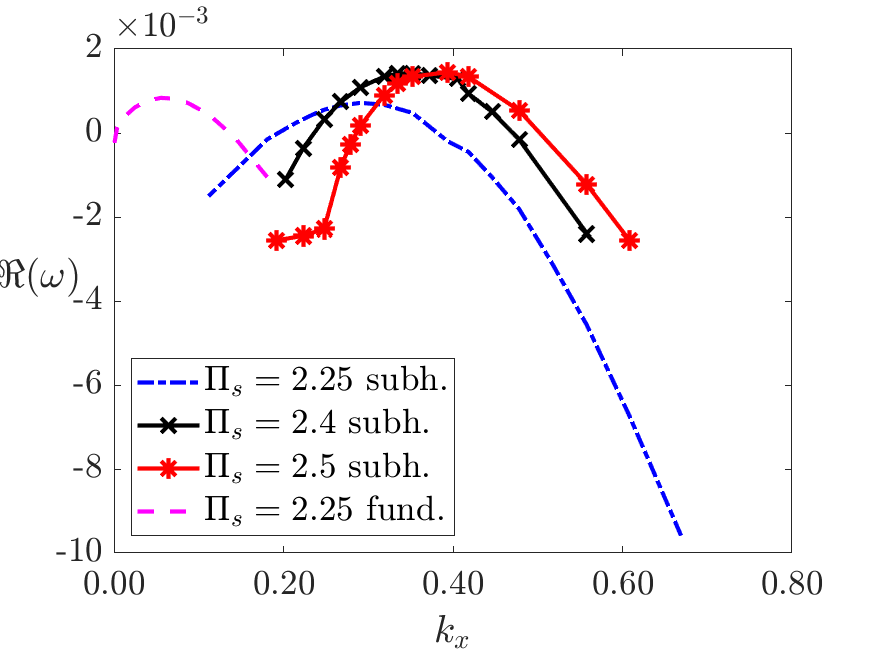}
   \caption{ }
	\label{subfig:subh08_gr}
\end{subfigure}
 \hfill
\begin{subfigure}{0.49\textwidth}
\centering
    \includegraphics[width=1.0\textwidth]{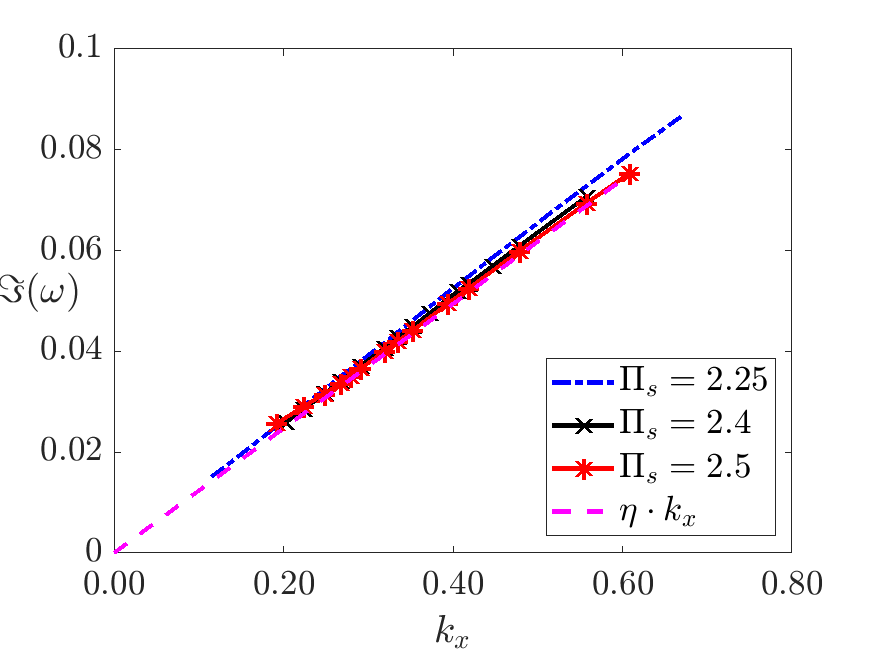}
   \caption{ }
	\label{subfig:subh08_osc}
\end{subfigure}
\vfill
\vspace{15pt}
 \begin{subfigure}{0.49\textwidth}
 \centering
 \vspace{2pt}
    \includegraphics[width=1.0\textwidth]{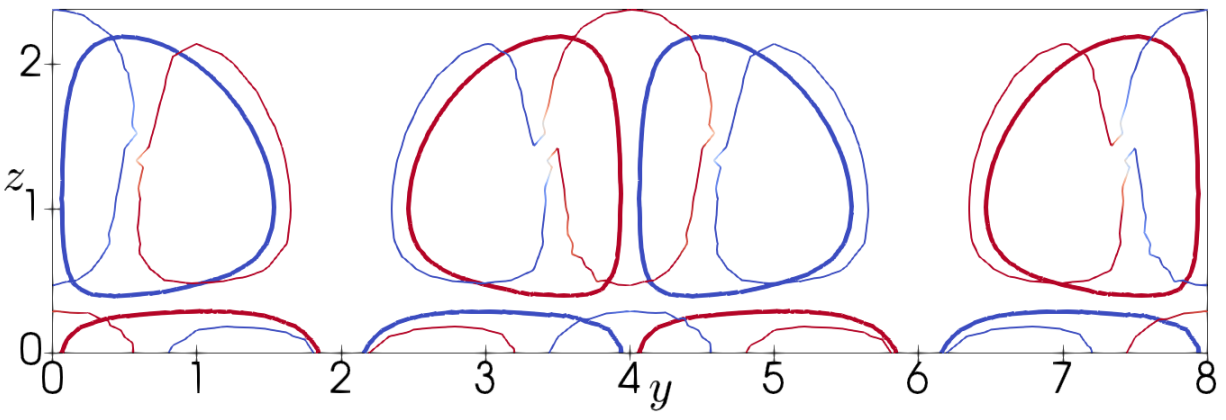}
   	\caption{}
	\label{subfig:subh08_2dcontour}
\end{subfigure}
\hfill
\hspace{10pt}
\begin{subfigure}{0.49\textwidth}
\centering
    \includegraphics[width=0.65\textwidth]{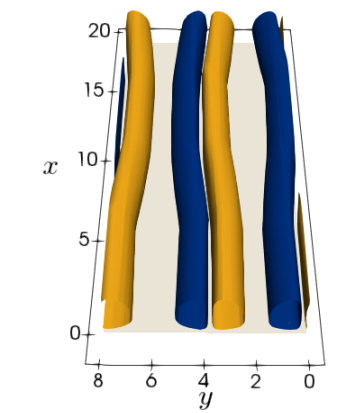}
   	\caption{}
	\label{subfig:subh08_3dcontour}
\end{subfigure}	
\caption{Characteristics of  the subharmonic mode for intermediately packed speaker-wire vortices  at different values of $\Uppi_s=2.25,2.4,2.5$ as a function of the longitudinal wave number $k_x$  : (a) Growth rates $\Re(\omega)$ of 3D modes along with the fundamental mode at $\Pi_s=2.4$ for comparison ;  (b) oscillation frequency $\Im(\omega)$ along with the linear dispersion relation $\eta\cdot k_x,\eta\approx 0.124$; (c) contours on the transverse $yz$ plane  of streamwise vorticity  for the 3D subharmonic instability(thin lines) at $\Pi_s=2.4$ in relation to the base flow(thick lines) and (d) streamwise vorticity contours of  base speaker-wire vortices perturbed by the subharmonic instability at $\Pi_s=2.4$.  \textcolor{blue}{All  contours are shown at 5\% of the maximal vorticity magnitude.}      }
\end{figure}

To uncover both subharmonic and fundamental  instabilities of the base speaker-wire vortices, we used two sets of  base flows for modal analysis consisting of two and one pair of speaker-wire vortices arising from the primary linear instability mode, respectively.  The  growth rates of the most unstable secondary modes for streamwise wave numbers  $k_x$ within the range $[0,0.8]$ are shown for three different values of $\Pi_s =2.25,2.4,2.5$ in figure \ref{subfig:subh08_gr}. In contrast to the previous configuration Case I with a closely spaced speaker-wire vortices, there are no unstable 2D modes, i.e. modes with a positive growth rate at $k_x=0$.
As expected, the maximal growth rate of the secondary  instability  is larger for the higher value $\Pi_s=2.4$ than for the smaller value $\Pi_s=2.25$, and the optimal streamwise wave number with maximal growth is also larger for $\Pi_s=2.4$. 
figure \ref{subfig:subh08_gr} also shows that the subharmonic mode for $\Pi_s=2.25$   attains its maximal growth rate at a larger streamwise wave number of $k_x\approx 0.4$ than its fundamental counterpart,  which is strongest at the  wave number $k_x\approx 0.1$, hence the unstable subharmonic modes tend to have a  clearly smaller streamwise wavelength than the fundamental ones.
A comparison between subharmonic and fundamental modes at the same value $\Pi_s=2.25$ also shows that in contrast to the base configuration with smaller vortex separation as described above, the subharmonic mode
does not achieve a larger maximal growth rate than its fundamental counterpart. This implies that in contrast to the previous base vortex configuration with a smaller vortex separation   $b_y$, the fundamental instability is of equal importance to the subharmonic mode. Thus, in addition to the merger between neighboring speaker-wire vortices due to the subharmonic mode, we would also  expect to see the effect of the fundamental mode during flow transition,  which could be a parallel translation of all vortices or (anti-)symmetric bending of sister rolls as described by \cite{pierrehumbert1982vortexinstab,corcos1984mixing}.  
\textcolor{black}{Due to symmetry,  this transverse translation could move equally likely along either positive or negative $y$ direction, depending on the nature of the initial disturbance.}

We observe from figure \ref{subfig:subh08_osc} that the  normalised frequencies of the most dominant subharmonic modes at the three values for $\Pi_s$ shown here appear to fit a linear dispersion relation given by $\Im(\omega)=\eta\cdot k_x$, where  $\eta \approx 0.124$  is empirically determined to approximately match all three curves. The accuracy of this linear relation demonstrates that the group speed  of the strongest subharmonic vortex instabilities given by $c=\frac{\partial \Im(\omega)}{\partial k_x}=\eta$ is nearly constant  and equals $\eta$. It should be noticed that this value for $\eta\approx0.124$ is roughly three times as small as its  value for Case I with $\eta\approx 0.327$. This means that the subharmonic instability waves travel far more slowly on base vortices with an intermediate vortex separation  than on more closely packed speaker-wire vortices.

As shown in  figure \ref{subfig:subh08_2dcontour},  the transverse wavelength of the subharmonic mode is  approximately \textcolor{black}{the transverse domain length}, hence  twice the base speaker-wire vortex separation $b_y$.
figure \ref{subfig:subh08_3dcontour} shows that
three-dimensional secondary subharmonic modes   bend  both sister rolls within the same speaker-wire vortex in the same direction, but pull neighboring rolls in 
adjacent speaker-wire vortices in opposite directions. Thus, the reconnection and merger  between rolls from different speaker-wire vortices is facilitated over a merger within a single speaker-wire vortex. \\

\begin{figure}
 \begin{subfigure}{0.48\textwidth}

	\centering
	\includegraphics[width=1\textwidth]{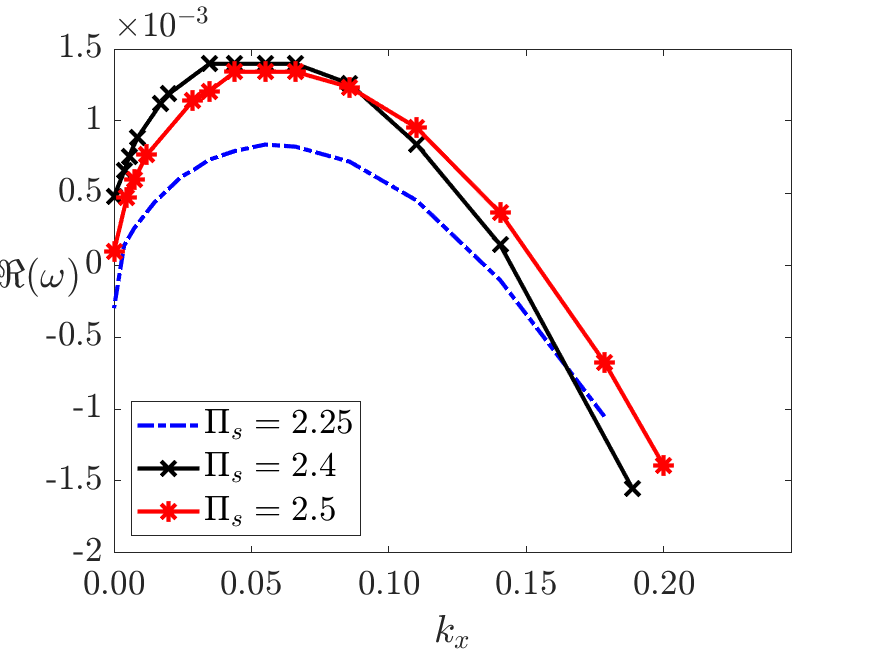}
   \caption{ }
	\label{subfig:fund04_gr}
 \end{subfigure}
  \hfill
  \begin{subfigure}{0.48\textwidth}
	\centering
	\includegraphics[width=1\textwidth]{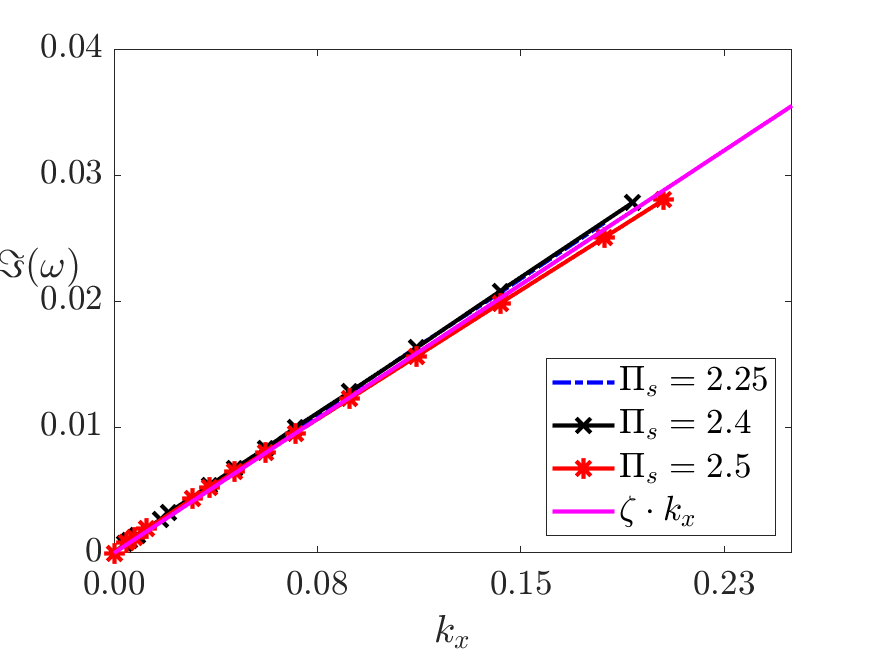}
   \caption{ }
	\label{subfig:fund04_osc}
 \end{subfigure}
 \vfill
 \vspace{15pt}
   \begin{subfigure}{0.48\textwidth}
\centering
	\includegraphics[width=0.95\textwidth]{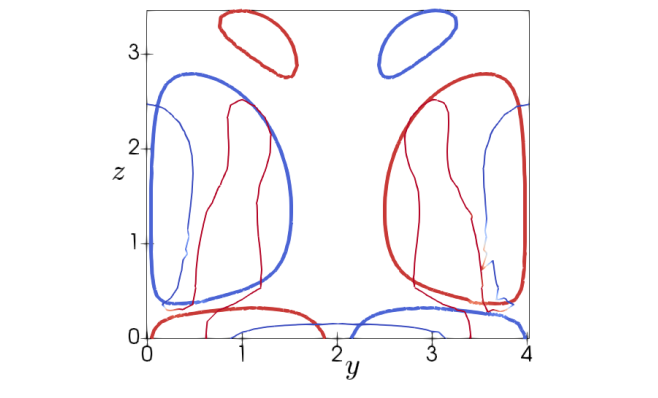}
   	\caption{}
	\label{subfig:fund04_contour2d}
 \end{subfigure}
  \hfill
      \hspace{35pt}
    \begin{subfigure}{0.45\textwidth}
	\includegraphics[width=0.6\textwidth]{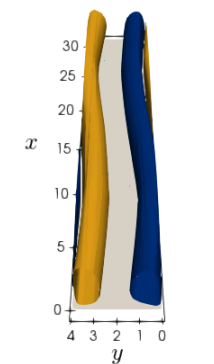}
   	\caption{}
	\label{subfig:fund04_contour3d}
 \end{subfigure}
 
	\caption{ Characteristics of  the fundamental mode for intermediately packed speaker-wire vortices   at different values of $\Uppi_s$ as a function of the longitudinal wave number $k_x$  : (a) Growth rate $\Re(\omega)$;  (b) oscillation frequency $\Im(\omega)$ along with the linear relation $\zeta\cdot k_x,\zeta=0.142$; (c)  streamwise vorticity contours  of two  adjacent rolls from  neighboring speaker-wire vortices on transverse plane for the 3D fundamental instability at $\Pi_s=2.4$  and (d)  streamwise vorticity contours of two rolls from neighboring  speaker-wire vortices perturbed by the fundamental instability at $\Pi_s=2.4$. \textcolor{blue}{All shown contours are at 5\% of the maximal vorticity magnitude.}  }
\end{figure} 

The growth rates and oscillation frequencies of the strongest fundamental modes on base vortices with intermediate vortex separation  are displayed in figure \ref{subfig:fund04_gr} and Figure \ref{subfig:fund04_osc}. They confirm the previous observation that the fundamental instability is a long-wave mode, which attains its maximum growth rate at wave streamwise numbers $k_x<0.1$ compared to the optimal frequency of the subharmonic mode at around $k_x\approx 0.4$. Figure \ref{subfig:subh08_osc} shows that the  oscillation frequencies of the most dominant fundamental modes at the three different values for $\Pi_s$ shown here seem to  follow the linear dispersion relation given by $\Im(\omega)=\zeta\cdot k_x$ with  $\zeta \approx 0.142$ being slightly larger than the value $\eta\approx 0.124$ for the subharmonic modes.  This value for $\eta\approx0.124$ is roughly one half of its  value at the lesser base vortex separation   discussed above with $\eta\approx 0.25$, which indicates that the fundamental modes  travel half as fast on base vortices with an intermediate  vortex separation than on more densely packed base vortices.

As   figure \ref{subfig:fund04_contour2d} shows,  the transverse wavelength of the fundamental mode is  approximately 4 and  equals the base speaker-wire vortex separation $b_y$.
figure \ref{subfig:fund04_contour3d} indicates that the
three-dimensional  fundamental mode   bends all rolls within a speaker-wire vortex  in the same direction. Thus, the mode cannot generate  a  merger or reconnection of the two rolls within a single speaker-wire vortex. The effect of the two-dimensional fundamental mode is to cause a parallel translation of all rolls similar to  fundamental Stuart vortex instabilities described in \cite{pierrehumbert1982vortexinstab}, and will not be explicitly displayed here. \\

\subsubsection{Case III: Loosely packed speaker-wire vortices}

\noindent\textbf{Subharmonic and fundamental modes:} For an even larger speaker-wire vortex separation  $b_y =5.0$,  figure \ref{subfig:vort_a} shows that the VPR  decreases further to around 50\%. This indicates that beyond the linear growth phase of the primary instability, the respective centers of the two vortices within each speaker-wire  pair are moving further towards each other such that the width of each pair vortex is smaller than one half the distance $b_y$ between adjacent speaker-wire vortices. The implication of this tightening of each speaker-wire vortex is that there is now sufficient space between two pairs to fit in an additional speaker-wire vortex. 

\begin{figure}
 \begin{subfigure}{0.48\textwidth}

	\centering
	\includegraphics[width=1\textwidth]{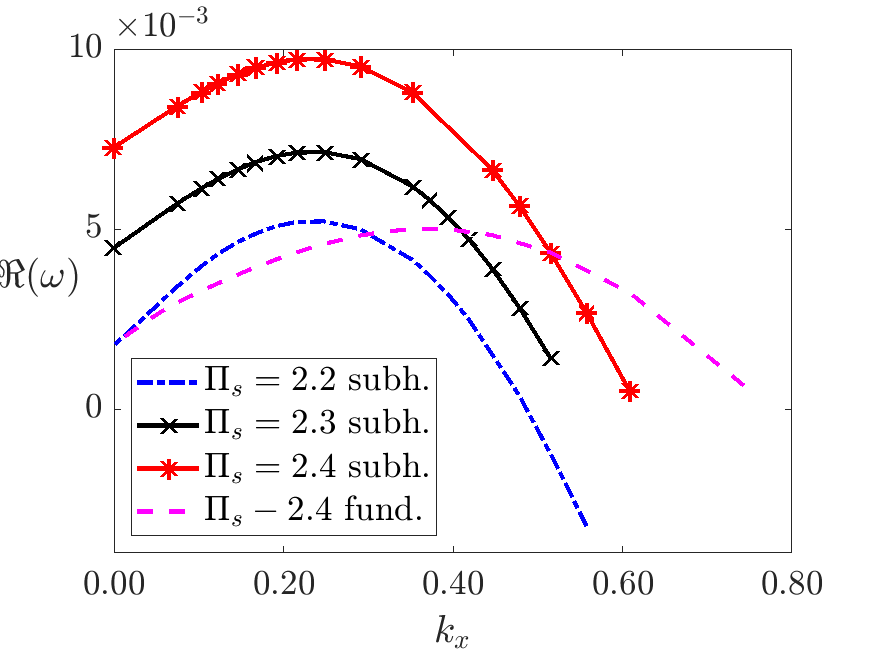}
   \caption{ }
	\label{subfig:subh105_gr}
 \end{subfigure}
  \hfill
  \begin{subfigure}{0.48\textwidth}
	\centering
	\includegraphics[width=0.99\textwidth]{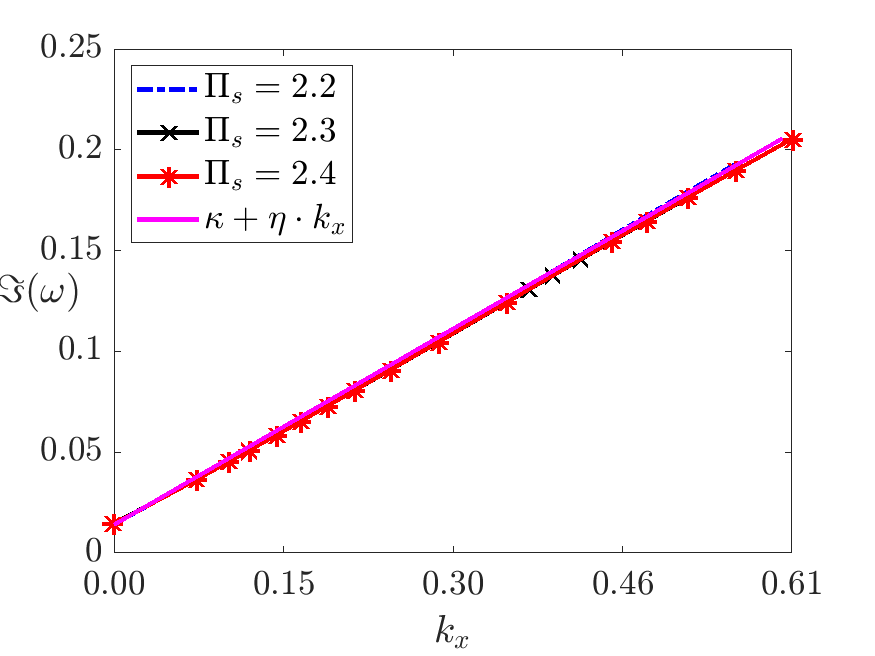}
   \caption{ }
	\label{subfig:subh105_osc}
 \end{subfigure}
 \vfill
 \begin{subfigure}{0.49\textwidth}
	\includegraphics[width=0.99\textwidth]{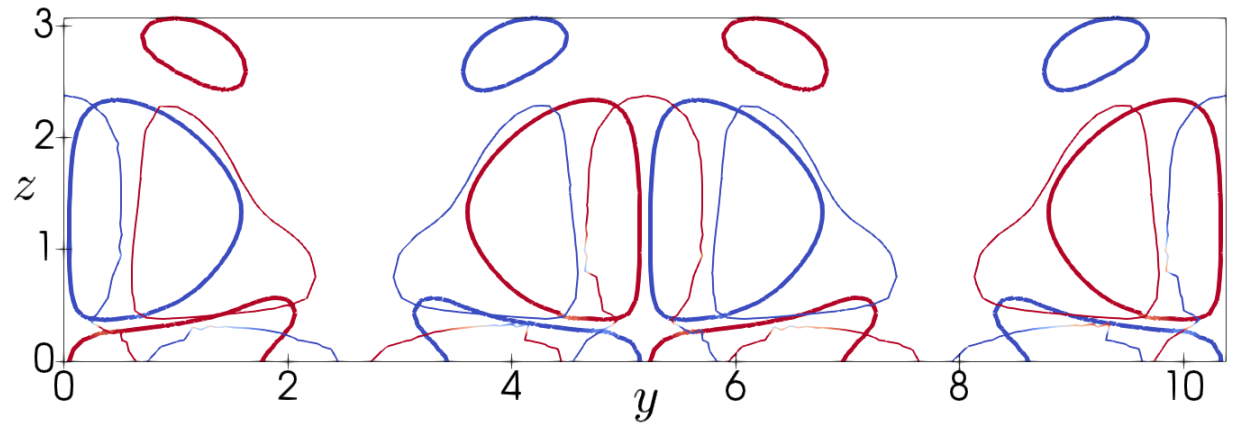}
   	\caption{}
	\label{subfig:subh105_2dcontour}
 \end{subfigure}
  \hfill
  \centering
  \begin{subfigure}{0.5\textwidth}
	\includegraphics[width=0.8\textwidth]{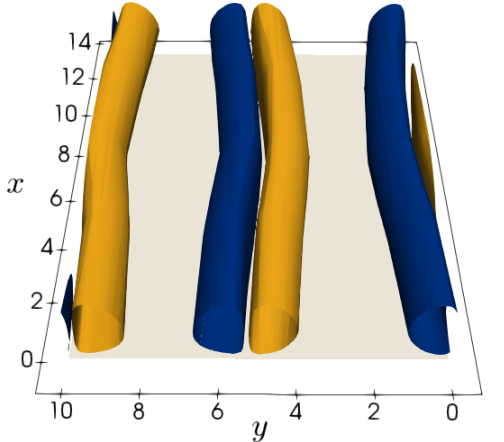}
   	\caption{}
	\label{subfig:subh105_3dcontour}
 \end{subfigure}
 
 \centering
  \begin{subfigure}{0.5\textwidth}
	\includegraphics[width=0.8\textwidth]{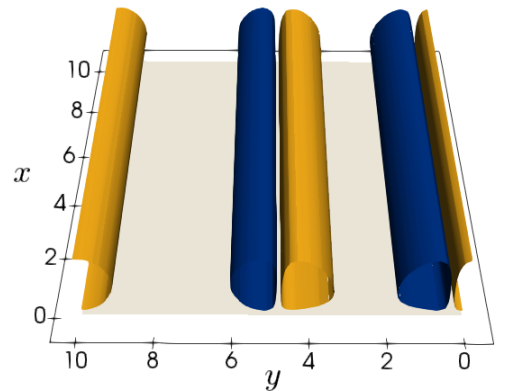}
   	\caption{}
	\label{subfig:subh105_3dcontour2d}
 \end{subfigure}
 
	\caption{ Characteristics of  the subharmonic mode for loosely packed speaker-wire vortices  at different values of $\Uppi_s$ as a function of the longitudinal wave number $k_x$  : (a) Growth rate $\Re(\omega)$: (b) oscillation frequency $\Im(\omega)$ with the linear relation $\kappa+\eta\cdot k_x, \kappa=0.014,\eta=0.32 $; (c) contours of streamwise vorticity on transverse plane  for the subharmonic instability mode (thin lines) at $\Pi_s=2.4$ in relation to the base flow(thick lines); (d)   streamwise vorticity contours of  base speaker-wire vortices perturbed by the subharmonic instability at $\Pi_s=2.4$ 	and (e) streamwise vorticity contours of  base speaker-wire vortices perturbed by two-dimensional subharmonic instability.  Videos of the evolution of the three-dimensional and two-dimensional subharmonic instabilities for the loosely packed vortices are available in the supplementary material as Movie 2 and Movie 3. \textcolor{blue}{All  contours are displayed at 5\% of the maximal vorticity magnitude.}  }
\end{figure}

To compute subharmonic and fundamental  instabilities for this base speaker-wire vortexconfiguration, we used  two sets of  base flows for modal analysis:  Two  speaker-wire vortices for the subharmonic mode and  one single   pair for the fundamental mode, all with the pair spacing  $b_y=5.0$.%
The  growth rates of the most unstable secondary modes for streamwise wave numbers  $k_x$ within  $[0,0.8]$ are shown for  different values of $\Pi_s =2.2,2.3,2.4$ in figure \ref{subfig:subh105_gr}, which indicates that the most dominant modes are subharmonic.
Same as for the other cases, the maximal growth rate of the secondary  instability  is larger for the higher value $\Pi_s=2.4$ than for the smaller value $\Pi_s=2.3$, and the optimal streamwise wave number $k_x$ with maximal growth rate is also larger for   $\Pi_s=2.4$ than for $Pi_s=2.3$. 

\begin{figure}
 \begin{subfigure}{0.48\textwidth}

	\centering
	\includegraphics[width=1\textwidth]{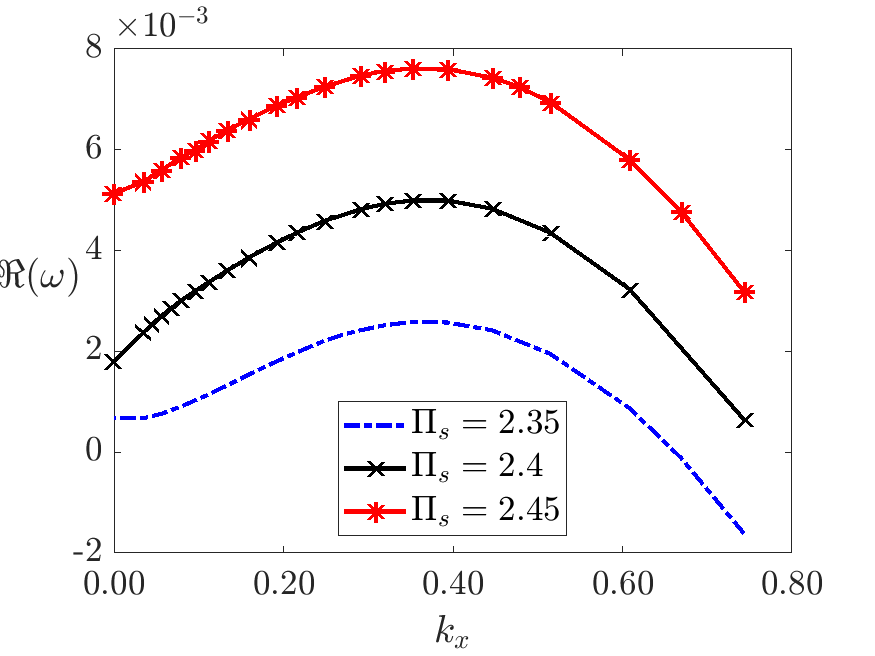}
   \caption{ }
	\label{subfig:fund05_gr}
 \end{subfigure}
  \hfill
  \begin{subfigure}{0.48\textwidth}
	\centering
	\includegraphics[width=1\textwidth]{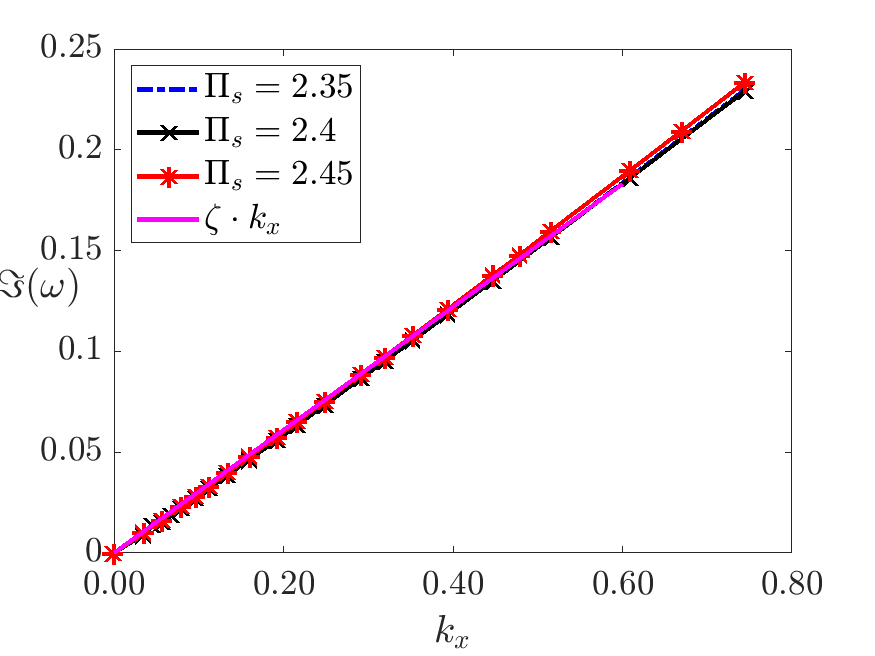}
   \caption{ }
	\label{subfig:fund05_osc}
 \end{subfigure}
  \vfill
 \vspace{15pt}
 \begin{subfigure}{0.49\textwidth}
\centering
	\includegraphics[width=0.99\textwidth]{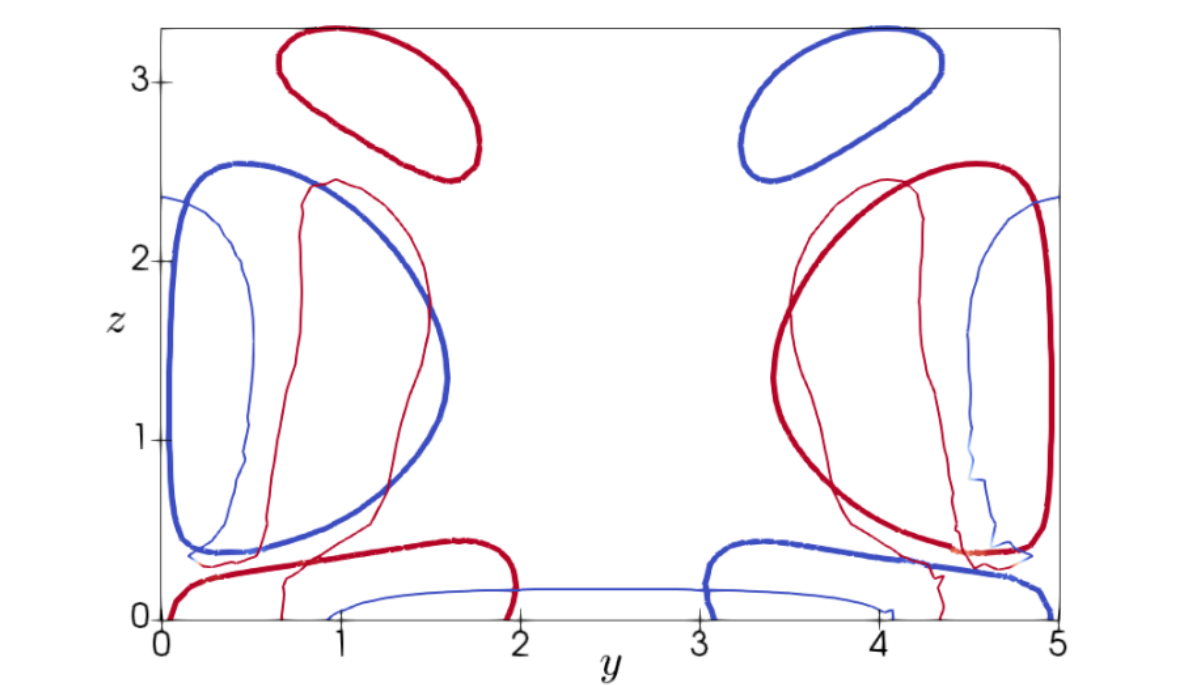}
   	\caption{}
	\label{subfig:fund05_contour2d}
 \end{subfigure}
  \hfill
     \begin{subfigure}{0.4\textwidth}
	\includegraphics[width=0.8\textwidth]{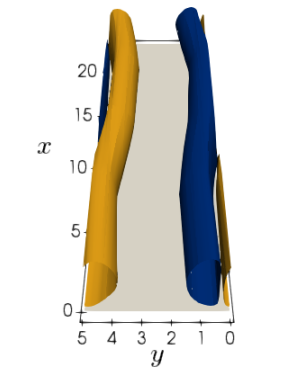}
   	\caption{}
	\label{subfig:fund05_3dcontour3d}
 \end{subfigure}
 
	\caption{ Characteristics of  the fundamental mode for  loosely packed speaker-wire vortices  at different values of $\Uppi_s$ as a function of the longitudinal wave number $k_x$  : (a) Growth rate $\Re(\omega)$;  (b) oscillation frequency $\Im(\omega)$ and the linear dispersion relation $\Im(\omega)=\zeta\cdot k_x,\zeta=0.305 $; (c) streamwise vorticity  contours of  two adjacent  rolls from neighboring speaker-wire vortices on transverse $yz$ plane for the fundamental instability(thin lines) at $\Pi_s=2.4$  and (d)  streamwise vorticity contours of  base speaker-wire vortices perturbed by the fundamental instability at $\Pi_s=2.4$. Videos of the evolution of the three-dimensional and two-dimensional fundamental instabilities for the loosely packed vortices are available in the supplementary material as Movie 4 and Movie 5. \textcolor{blue}{All  contours are shown at 5\% of the maximal vorticity magnitude.} 
	}
\end{figure}

Comparison with the growth rate of the 3D fundamental mode at $\Pi_s= 2.4$ in figure \ref{subfig:subh105_gr} shows that the subharmonic mode at the same $\Pi_s$ value achieves a clearly larger growth rate for all streamwise wave numbers $k_x$. This assertion is further supported when we compare the growth rates for fundamental modes at $\Pi_s=2.35,2.4,2.45$ shown in figure \ref{subfig:fund05_gr}  against the growth rates shown in figure \ref{subfig:subh105_gr} for subharmonic modes. This means that similar to Case I with the smallest vortex separation ($b_y=3.1$), but in contrast to Case II with an intermediate spacing of the speaker-wire vortices ($b_y =4.1$), there exists a gap between the subharmonic and fundamental modes when all other base flow parameters are the same. Thus, we would   expect to see the dominant effect of the subharmonic mode at $\Pi_s \approx 2.4$ during flow transition. However, as will be shown later, this behavior can change at larger $\Pi_s$ values.  


The  oscillation frequencies of the strongest subharmonic modes for base vortices with large vortex separation are displayed in figure \ref{subfig:subh105_osc}, showing that the   frequencies of the most dominant fundamental modes at the three different values for $\Pi_s$ displayed here appear to  fit a linear dispersion relation given by $\Im(\omega)=\kappa+\eta\cdot k_x$ with  $\kappa \approx 0.014, \eta\approx 0.32$.
This shows that, in contrast to the previous configurations with smaller base  vortex separations,  even the two-dimensional subharmonic instability ($k_x=0$) is oscillatory with frequency $\kappa$.
The group velocity  $\eta$ for the subharmonic modes  is approximately the same as the corresponding value for subharmonic modes at small base  vortex separation which is $\eta\approx 0.327$ and  clearly larger than the value at intermediate base  vortex separation. 

As shown in  figure \ref{subfig:subh105_2dcontour},  the transverse wavelength of the subharmonic mode is  approximately 10, hence  twice the value of the base speaker-wire vortex separation $b_y$.
figure \ref{subfig:subh105_3dcontour} shows that
three-dimensional  subharmonic modes   bend  both sister rolls within the same speaker-wire vortex in the same direction, but pull neighboring rolls in 
adjacent speaker-wire vortices in opposite directions. Thus the reconnection and merger  between rolls from different speaker-wire vortices is facilitated over a merger within a single speaker-wire vortex. The effect of the two-dimensional subharmonic mode is displayed in figure \ref{subfig:subh105_3dcontour2d}: It can be seen that one pair of neighboring speaker-wire vortices move closer toward each other  in the precursor to a merger, while at the same time distancing themselves from their other neighbors and increase the in-between space where additional vortices can be formed.
This can be seen in the animations of the evolution of subharmonic instabilities obtained via \textcolor{black}{numerical integration of the three-dimensional Navier-Stokes  Equations (\ref{eqnslopemom})-(\ref{eqnslopebuoy}) }  available as supplementary  Movie 2 and Movie 3. 
\textcolor{blue}{As displayed in the attached movies, the nonlinear dynamics subharmonic instability for
well-separated base vortices are different than a mere merger which occurs for vortices that are more closely packed. As predicted by linear stability analysis, the subharmonic mode manifests itself initially by bending neighboring pairs in opposite directions. However, at later times, novel vortices emerge in the free space between the pairs, and those newly created vortices will later merge with one of the original base vortices. When the spacing between base vortices is small, then this phenomenon of vortex creation cannot occur due to the lack of free spacing, hence the only dynamics is the merger of adjacent speaker-wire vortices. This supports our main message that the vortex instability dynamics is strongly dependent on the spacing between adjacent pairs, which is quantified by the vortex packing ratio(VPR) introduced in the text. However, linear stability analysis by itself can only predict the initial onset of vortex instability, but the subsequent nonlinear evolution of the vortices as shown in the movies.}
\\

The growth rates and oscillation frequencies for the strongest fundamental modes on loosely packed base vortices with large vortex separation are displayed in figure \ref{subfig:fund05_gr} and Figure \ref{subfig:fund05_osc}.  In contrast to base vortices with intermediate vortex separation discussed above, the fundamental instability  attains its maximum growth rate at higher wave streamwise numbers $k_x\approx 0.4$ compared to the most dangerous frequency of the subharmonic mode at around $k_x\approx 0.2$. As shown in figure \ref{subfig:fund05_osc}, the   frequencies of the strongest fundamental modes at the three different values for $\Pi_s$ displayed here closely  follow the linear dispersion relation given by $\Im(\omega)=\zeta\cdot k_x$ with  $\zeta \approx 0.305$, which is only  slightly smaller than the value $\eta\approx 0.32$ for the subharmonic modes.  This would indicate that the fundamental modes  travel at around the same speed as their subharmonic counterparts.

As displayed in  figure \ref{subfig:fund05_contour2d},  the transverse wavelength of the fundamental mode is  approximately 5, which  equals the base speaker-wire vortex separation $b_y$.
figure \ref{subfig:fund05_3dcontour3d} indicates that
three-dimensional  fundamental instabilities   bend all rolls within speaker-wire vortices  along the same direction. Thus, they by themselves cannot generate  the dynamics for  merger within a single speaker-wire vortex. The effect of the two-dimensional fundamental mode is to cause a parallel translation of all rolls similarly as  fundamental Stuart vortex instabilities described in \cite{pierrehumbert1982vortexinstab}, and will not be explicitly displayed here. \\

\begin{figure}
 \begin{subfigure}{0.48\textwidth}

	\centering
	\includegraphics[width=0.92\textwidth]{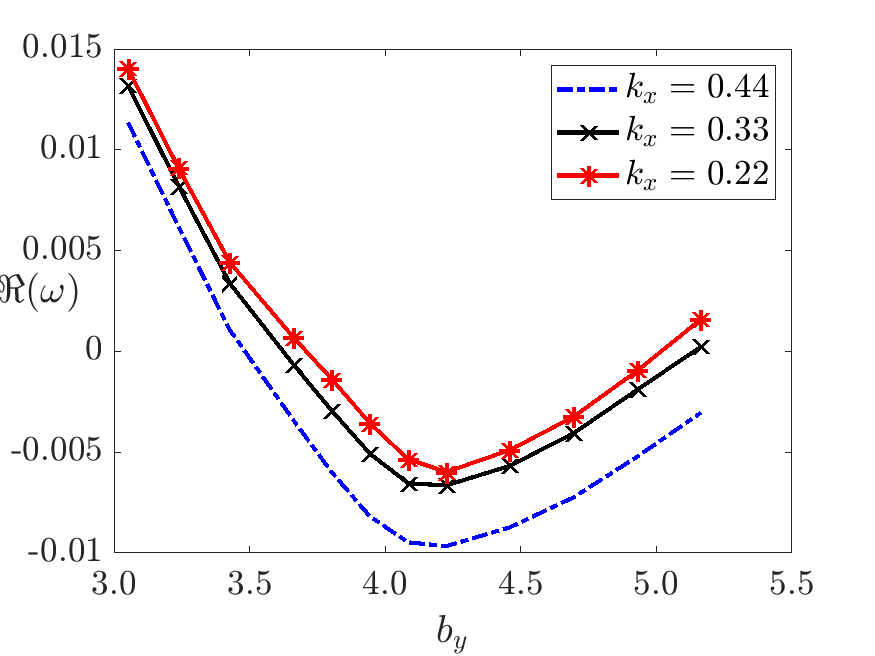}
   \caption{ }
	\label{subfig:lambda_subh_gr}
 \end{subfigure}
  \begin{subfigure}{0.48\textwidth}
	\centering
	\includegraphics[width=0.92\textwidth]{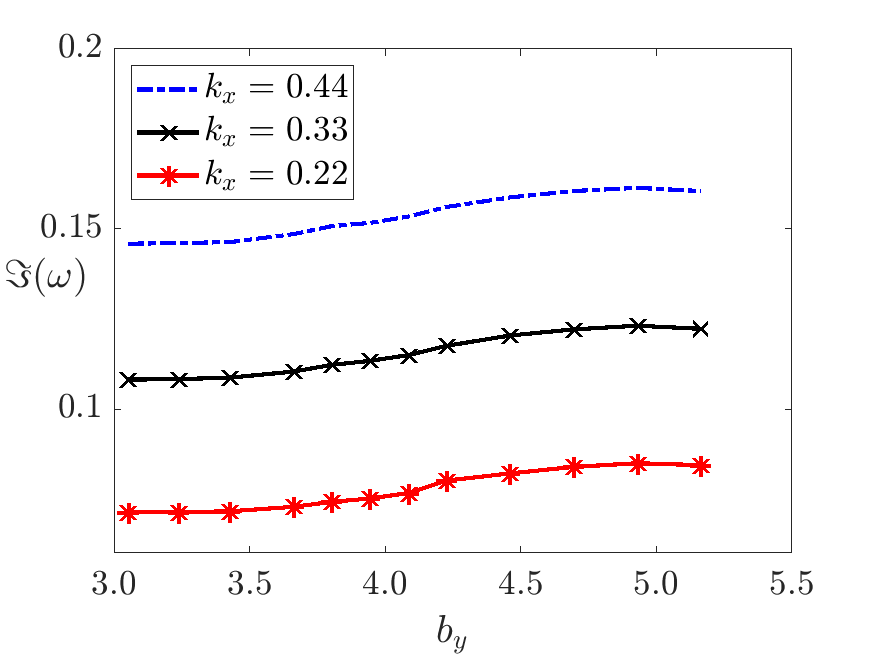}
   \caption{ }
	\label{subfig:lambda_subh_osc}
 \end{subfigure}
 
	\caption{ Characteristics of subharmonic secondary instability for different streamwise wave numbers $k_x$ at  $\Pi_s=1.9$  as a function of the  vortex separation $b_y$ (which is also the transverse wavelength of the base vortices):  (a) Growth rate and  (b) oscillation frequency }\label{fig:anasubh_lambda}
\end{figure} 

\begin{figure}
 \begin{subfigure}{0.48\textwidth}

	\centering
	\includegraphics[width=0.99\textwidth]{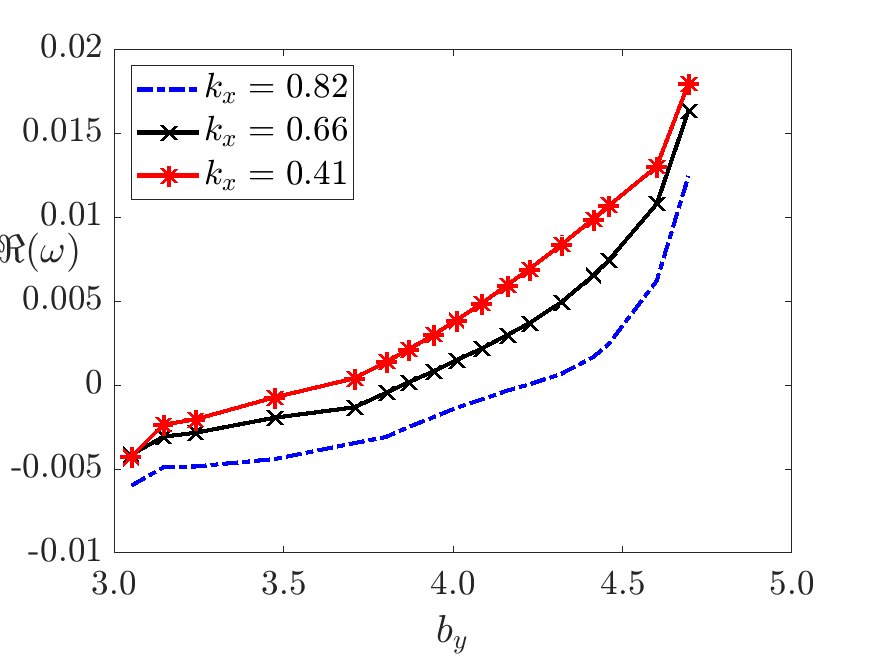}
   \caption{ }
	\label{subfig:lambda_fund_gr}
 \end{subfigure}
  \begin{subfigure}{0.48\textwidth}
	\centering
	\includegraphics[width=0.99\textwidth]{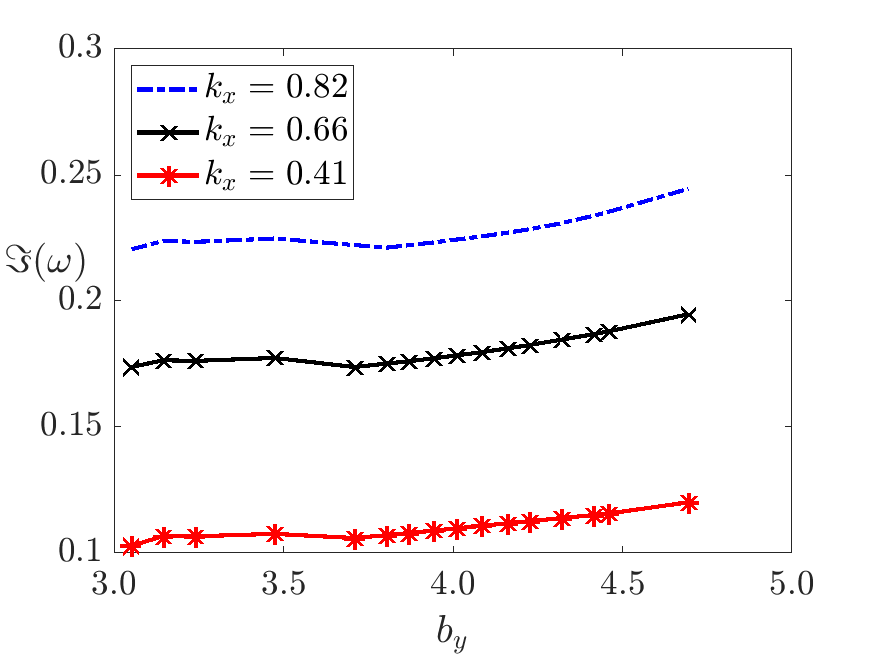}
   \caption{ }
	\label{subfig:lambda_fund_osc}
 \end{subfigure}
 
	\caption{ Characteristics of the fundamental secondary instability for different streamwise wave numbers $k_x$ at  $\Pi_s=2.85$  as a function of the  base  vortex separation $b_y$ :  (a) Growth rate compared to primary linear instability and  (b) oscillation frequency }\label{fig:anafundh_lambda}
\end{figure} 

\subsection{Effect of speaker-wire vortex spacing on instability }

As we have seen from the previous presentations of instabilities for different speaker-wire vortexconfigurations, the  distance $b_y$ between adjacent speaker-wire vortices is crucial in determining which kind of modes are favored to destabilize the vortices.
For both small and large values of $b_y$, our prior analysis suggests that subharmonic modes, i.e. those with transverse wavelength $\lambda_y=2b_y$, are clearly stronger than their fundamental counterparts with half the wavelength $\lambda_y=b_y$. However, when $b_y$ takes on an intermediate value, both fundamental and subharmonic modes may have similar growth rates at smaller stratification numbers $\Pi_s$. 

To investigate the influence of  vortex separation $b_y$ on secondary instability properties in a more systematic way, we have prepared a series of base flows containing speaker-wire vortices with vortex separation $b_y$ within the range  $[3.0, ~ 5.2]$ and computed their most unstable subharmonic as well as fundamental modes. For the study of subharmonic modes, we chose the stratification number $\Pi_s=1.9$ for all base flow configurations which only differ by their  vortex separation $b_y$. The growth rates and oscillation frequencies of the most unstable 3D subharmonic modes with different streamwise wave numbers $k_x=0.22,0.33,0.44$ as a function of $b_y$ are shown in figure \ref{subfig:lambda_subh_gr} and figure \ref{subfig:lambda_subh_osc}. These results corroborate our earlier observations about the dependence of subharmonic mode  on  vortex separation $b_y$: The subharmonic instability assumes its strongest growth rate when $b_y$ is the smallest and becomes weaker when $b_y$ is increased until reaching its minimum, which is negative, at an intermediate value of $b_y\approx 4.2$. From that point onwards, the subharmonic instability starts to get stronger again with further increases in $b_y$; however, at the largest $b_y=5.1$ shown here, even though again at a positive growth rate, it never regains its peak strength at the smallest vortex separation $b_y=3.1$. Figure \ref{subfig:lambda_subh_osc} shows that in contrast to the growth rate, the oscillation frequency of the strongest subharmonic mode at a given wave number $k_x$ generally grows monotonically with vortex separation $b_y$  for all  $b_y\leq 4.5$. For larger vortex separation values $b_y$, the oscillation frequency visibly stagnates at its value for the vortex separation $b_y=4.5$.

To study the fundamental instability modes, we have chosen the stratification number $\Pi_s=2.85$ for all base flow configurations. The growth rates and oscillation frequencies of the most unstable 3D fundamental modes at three different streamwise wavnumbers $k_x=0.41,0.66,0.84$ as a function of $b_y$ are shown in figure \ref{subfig:lambda_fund_gr} and figure \ref{subfig:lambda_fund_osc}. These results show that in contrast to the subharmonic modes, the fundamental instability  increases in strength with growing vortex separation $b_y$, starting with negative growth rates at $b_y=3.1$ and becoming positive for $b_y>4.0$. A physical explanation for this could be that at larger distances between adjacent speaker-wire vortices, the influence between vortices within a pair outweighs the effect of the more distant vortices at a different vortex pair. Figure \ref{subfig:lambda_fund_osc} shows that the oscillation frequency of the strongest fundamental mode also grows with increasing $b_y$ throughout the entire range as displayed here.

\section{Conclusions}

We have carried out a linear bi-global  stability analysis for stationary longitudinal rolls that emerge as a primary instability from the laminar one-dimensional anabatic Prandtl slope flow at low angles. We have identified the pair of counter-rotating rolls as unique flow structures and designated them as \textit{speaker-wire vortices}. Despite the apparent resemblance of the longitudinal rolls in our study with other well-known counter-rotating vortex pairs  reported in the literature, 
   our base flow configuration has  multiple distinctly unique features. Firstly, it includes an independent background stratification that is at an oblique angle to the heated solid slope surface.  Secondly,    the  rolls in our base flow configuration  have \textit{three} non-zero velocity components even though the flow field itself  has zero gradient along the streamwise direction. As a result, it is not surprising that the instability dynamics of these speaker-wire vortices are also distinct from the hitherto known vortex pair instabilities. 
   
   Our results for secondary instabilities of speaker-wire vortices have shown that the vortex packing ratio (VPR), \textcolor{black}{ or equivalently,} 
the  vortex separation $b_y$, i.e. the distance between two adjacent speaker-wire vortices, 
plays a major role in determining which stability mode is the most significant in destabilizing the vortices under anabatic slope flow conditions.
The growth rate of the most dominant subharmonic mode as a function of vortex spacing $b_y$ has a parabola-like shape: It is maximal at the lowest values $b_y\leq 3.1$, then gradually decreases to its minimum at $b_y\approx 4.4$ when the base vortex configuration is the most stable, but starts increasing again for larger $b_y$. In contrast, the growth rate of the strongest fundamental mode  increases with larger vortex separation $b_y$.
For closely-packed as well as sparsely-packed vortex configurations, i.e. those with VPR over 70\% or less than 50\%, 
the subharmonic secondary instability is far more dominant than the fundamental modes, which only become unstable at higher normalised surface heat flux values.   On the other hand, when VPR drops to 60\% at an intermediate  vortex separation $b_y\approx 4.1$, the long-wave fundamental mode may become roughly equal to the strongest subharmonic mode. 
For all the subharmonic and fundamental instability modes computed for speaker-wire vortices of different VPR and vortex separation values, we have found that rolls within the same speaker-wire vortex always bend in the same direction, hence keeping the same distance toward each other even after being destabilised. 
The implication of this dynamic is that in  speaker-wire vortices,   reconnections are only possible when at least two speaker-wire vortices  (i.e. four vortex rolls) are present, whereas  two rolls  within a single speaker-wire vortex are unlikely to reconnect with each other even after the onset of instabilities. This dynamic is in contrast to most other well-known vortex instabilities such as the Crow-instability, the elliptic instability or the zig-zag instability, which has led us to identify the speaker-wire vortex as a novel coherent vortex unit. 
  
 
The subharmonic and fundamental modes in  speaker-wire vortices come in both three-dimensional as well as two-dimensional types. The two-dimensional subharmonic mode for closely packed base vortex configurations acts to weaken every second vortex, eventually destroying them and freeing up vortex-free space between the remaining rolls. On the other hand, for sparsely-packed base vortex configurations, the two-dimensional subharmonic mode moves two adjacent speaker-wire vortices closer toward each other as a precursor to a vortex merger, which simultaneously  creates additional space between two other vortices in which additional vortices can be formed,  resulting in a more densely packed vortex array. 
The physical interpretation of this dynamic would be that these two-dimensional instabilities aim at generating a vortex configuration which has an optimal intermediate distance between each speaker-wire vortex and is neither too densely nor too sparsely packed. 
 The two-dimensional fundamental mode, which can only play a meaningful  role at an intermediate base vortex distance of $b_y\approx 4.1$ by not being dominated by its subharmonic counterpart, has the net effect of  displacing all base speaker-wire vortices along the transverse direction at the same speed. 
 
 
The three-dimensional secondary instability modes, i.e. those which vary along the streamwise or equivalently the along-slope direction, have been shown to cause streamwise sinusoidal bending of each vortex in the base configuration. The three-dimensional subharmonic mode, whose transverse wavelength is twice the base  vortex separation, moves two adjacent speaker-wire vortices closer to each other, and also bends the two neighboring rolls belonging to different speaker-wire vortices along opposite directions. As a result, the segments on each roll which are bent towards their neighbor from the adjacent speaker-wire vortex have the smallest distance to the corresponding segment in its neighboring vortex, eventually getting close enough to become sites of a vortex  reconnection and thus ripped from its sister roll in the original speaker-wire vortex. For tightly packed base vortex configurations, the growth rates of three-dimensional subharmonic modes remain nearly constant for all  streamwise wave numbers smaller than that of the most dominant mode, thus revealing the presence of potential long-wave instabilities.
The three-dimensional fundamental mode, with a transverse wavelength equivalent to the vortex separation of the base speaker-wire vortices, bends all  rolls in all speaker-wire vortices along the same direction, thus the distances between  them remain the same as in the original base configuration before the onset of instability. Thus, the fundamental mode by itself cannot be the mechanism leading to vortex reconnection. Instead, after a sufficiently long time, the subharmonic instability
will eventually manifest itself by either moving two neighboring speaker-wire vortices towards each other (densely packed case) or creating  new vortices between two pairs (thinly packed case), resulting in the reconnection between two adjacent rolls which are not from the same speaker-wire vortex as in the base configuration as described above.

 One of the most poignant findings uncovered in our study is that for all fundamental and subharmonic speaker-wire vortex instabilities, the sister rolls within a single speaker-wire vortex always bend in the same direction, thus keeping the distance from each other and preventing  reconnection between them.
 This implies that the only possible  vortex reconnection dynamics under Prandtl's anabatic slope flow model requires  four vortex rolls in two speaker-wire vortices, whereas one single speaker-wire vortex is able to maintain its two-roll structure even after the onset of vortex instabilities. The sensitivity of vortex dynamics with respect to the base  vortex separation, which characterizes the degree of vortex array density along the transverse direction, is also a distinct feature not observed in other secondary instabilities such as those occurring in convective boundary layers
 Hence, the \textit{speaker-wire} vortex system described in this work merits its designation as  a novel coherent flow structure whose dynamics may have profound impact on turbulent transitions in stably stratified boundary layers.\\

\textbf{Supplementary movies.} Supplementary movies are available at https://.\\

\textbf{Acknowledgments:}
This material is based upon work supported by the National Science Foundation under Grant No. (1936445). Research was sponsored in part by the University of Pittsburgh, Center for Research Computing through the computing resources provided.

\textbf{Declaration of Interests}: The authors report no conflict of interest.

\bibliographystyle{jfm}
\bibliography{stability}

\end{document}